\documentclass[twocolumn,showpacs,superscriptaddress,nofootinbib,fleqn]{iopart}

\usepackage{amsfonts}
\usepackage{amssymb}
\usepackage{amstext}
\usepackage{bm}
\usepackage{cite}
\usepackage{dcolumn}
\usepackage{graphicx}
\usepackage{graphics}
\usepackage[latin1]{inputenc}
\usepackage{latexsym}
\usepackage{rotating}
\usepackage{url}
\usepackage{xspace} 
\usepackage{xcolor}
\usepackage{booktabs}
\usepackage{mathrsfs}
\usepackage[colorlinks=true]{hyperref}
\usepackage[bottom]{footmisc}
\usepackage[font=normalsize,labelfont=bf,justification=raggedright]{caption}
\usepackage{epstopdf}
\usepackage{gensymb}
\definecolor {darkgreen}{rgb}{0.2,0.7,0.2}

\newcommand\be{\begin{equation}}
\newcommand\ba{\begin{eqnarray}}
\newcommand\ee{\end{equation}}
\newcommand\ea{\end{eqnarray}}

\newcommand\bw{\begin{widetext}}
\newcommand\ew{\end{widetext}}

\newcommand{\hor}{{\mbox{\tiny H}}}

\newcommand{\sph}{{\mbox{\tiny sph}}}
\newcommand{\cir}{{\mbox{\tiny circ}}}

\newcommand{\scr}{{\mbox{\tiny scr}}}
\newcommand{\BH}{{\mbox{\tiny BH}}}
\newcommand{\BHS}{{\mbox{\tiny BHS}}}
\newcommand{\OS}{{\mbox{\tiny OS}}}

\begin{document}

\title{Observing the Shadows of Stellar-Mass Black Holes with Binary Companions}

\author{%
Harrison~Gott$^{1}$,
Dimitry~Ayzenberg$^{1,2}$,
Nicol\'as~Yunes$^{1}$,
and
Anne~Lohfink$^{1}$
}

\address{$^{1}$~eXtreme Gravity Institute, Department of Physics, Montana State University, Bozeman, MT 59717, USA.}
\address{$^{2}$~Center for Field Theory and Particle Physics and Department of Physics, Fudan University, 200438 Shanghai, China}

\date{\today}

\begin{abstract} 

The observation of the shadows cast by the event horizon of black holes on the light emitted in its neighborhood is the target of current very-long-baseline-interferometric observations. 
When considering supermassive black holes, the light source is the black hole's accretion disk, and therefore, the observation of the shadow may reveal information about the black hole and the accretion flow.
We here study the shadows cast by stellar-mass black holes that are illuminated not by an accretion disk but by a stellar companion in a wide binary orbit. 
We calculate the shadows produced in such a configuration for the first time and show snapshots of the time-dependent shadow ``movie'' that is generated. 
We also study the minimal criteria for detecting and resolving such shadows with very-long-baseline-interferometric observations. 
We find that one would need telescopes capable of resolving apparent magnitudes greater than $33$ with baselines larger than $10^{6}$ km. 
Therefore, although very-long-baseline-interferometric efforts such as the Event Horizon Telescope would not be able to detect these shadows, their detection is possible with future telescopes in the next few decades.

\end{abstract}

\maketitle

\section{Introduction}

In April 2017, the Event Horizon Telescope~\cite{2009astro2010S..68D} (EHT) collaboration undertook a week long observing campaign of Sagittarius A*, the supermassive black hole (BH) at the center of our Milky Way Galaxy. EHT, through the use of very long baseline interferometry techniques (VLBI), joins together telescopes around the world into a telescope that is effectively as large as the Earth. Thus, this literally world-wide telescope has an unprecedented observing resolution of about 20 $\mu$as, allowing it to resolve features on the scale of an orange on the Moon as seen from Earth. For EHT and similar projects such as BlackHoleCam~\cite{Goddi:2017pfy} and GRAVITY~\cite{2008SPIE.7013E..2AE}, the goal is to resolve features near to and on the scale of the event horizon of Sagittarius A* and other supermassive BHs. 

An observation of particular interest to EHT and similar projects is what is known as the BH shadow, a dark region that exists on any observation of a BH caused by photons falling into the BH event horizon. Unlike any other object in the universe, photon orbits exist around a BH. But these orbits are unstable and demarcate the boundary between photons that will escape the gravitational well of the BH and those that will be forever lost inside the event horizon. This boundary, the conglomeration of all the spherical unstable photon orbits, is known as the photon sphere and when observed far from the BH it is the boundary between the light and dark regions of the BH shadow image. The shape of the boundary depends on the properties of the BH and thus the shadow observation can be used to determine these properties.

Currently, only supermassive BHs are the targets of BH shadow observation campaigns. The radius of the photon sphere is on the order of 1 km for a BH the mass of the Sun and scales linearly with mass; thus, it is easy to see why supermassive BHs, with shadows on the order of $10^{6}$ km in radius, are the only current targets. Attempting to image the shadows of stellar-mass BHs, which have masses on the order of 10 $M_{\odot}$, and are typically found at distances of thousands of parsecs from Earth is impossible with current observational capabilities. Even so, with advances in technology and the construction of new telescopes, it may be possible in the coming future to achieve observations of shadows of stellar-mass BHs.

Generally, when discussing electromagnetic observations of BHs the primary source of the electromagnetic radiation near the BH is an accretion disk, i.e.~a disk of gas orbiting the BH. In the case of stellar-mass BHs, these disks form when gas from a companion star is pulled into an orbit around the BH. This most commonly occurs when the star overflows its Roche lobe, i.e.~the region within which the star's gravitational pull dominates. In more massive stars that are more widely separated, a disk can also form if the stellar wind of the star is large enough to send a significant amount of gas towards the BH. When the orbital separation is large enough and the stellar wind of the companion star is relatively small, an accretion disk will not form or will be scarce enough that the electromagnetic radiation of the companion star will dominate the spectrum. In this case, the primary source of illumination for the BH shadow observation will be the stellar companion, a scenario that has not been studied to date.

In this paper, we perform the first calculations of BH shadows produced by stellar-mass BHs in binary systems in which the companion star is the primary source of electromagnetic radiation. We calculate these shadows as follows: we employ a general relativistic ray-tracing code to solve for the trajectories of photons near a Kerr BH, and sub-select those that would have originated from a companion star, as a function of the physical properties of the binary. The end result are images of the BH shadow at different points in the orbit, which can, in principle, be combined to create a BH shadow movie. We find that the shadows are quite complex and can provide a wealth of information about the BH, the companion star, and the system as a whole, provided they are detected and a proper model is used to filter the data.

This paper is, in some sense, related to work done by Cunningham and Bardeen in 1973~\cite{1973ApJ...183..237C, Dokuchaev:2018fze}. In this previous work, the system modeled was that of a star in a very close (at most 10 gravitational radii) circular orbit around a supermassive extremal Kerr BH. The main results were the analytic calculation of the apparent position and energy flux of the star in different locations in its orbit. To perform these calculations the star was treated as a point source of electromagnetic radiation, although the leading order effects of an extended source were taken into account. Recently, Cunningham's and Bardeen's work was corrected and redone using the physical properties of Sagittarius A*~\cite{Dokuchaev:2018fze}, and the leading order effects on the image of the star were calculated and made into images and a movie. 

Our work also deals with the calculation of the image of a star orbiting a BH, but there are significant differences in the systems being studied and the method of calculation. We here focus on a binary system composed of a BH and a star, so the separation between the two bodies is several orders of magnitude larger than in previous work and the masses are within an order of magnitude. Our method for calculating the images is to solve the photon equations of motion numerically rather than using an analytic method. While our method introduces some numerical error, it is an exact calculation in the sense that we do not truncate the equations at a finite order. Due to these differences, our work and results are not directly comparable to the work of Cunningham and Bardeen~\cite{1973ApJ...183..237C, Dokuchaev:2018fze}.

In addition to calculating these shadows, we also discuss the feasibility of observing them with future instruments. Using some simplifying assumptions, we make rough estimates of the apparent magnitude and angular size of the shadow. We then evaluate these magnitudes for the three nearest BH-stellar binary systems~\cite{1986ApJ...308..110M, 2010ApJ...710.1127C, Ziolkowski:2005ag, 2011ApJ...742...84O, Parker:2015fja, 1994MNRAS.271L..10S, 1994MNRAS.271L...5C, Bernardini:2016mub} and the only non-interacting BH-stellar binary candidate~\cite{Thompson:2018ycv} and compare them to the capabilities of current telescopes. As expected, we find that current telescopes cannot detect or resolve the BH shadows of stellar-mass BHs, but we argue that given what is currently possible and what is planned for the very near future, it is not unreasonable to think that imaging these shadows will be possible in the next few decades. Such observations will require a VLBI telescope that is at least a few orders of magnitude more sensitive than the proposed Large UV/Optical/Infrared Surveyor~\cite{LUVOIR} (LUVOIR) and is placed at least $\sim10^{7}$ km from Earth. We suggest that the Lagrange points $L_{4}$ and $L_{5}$, located $\sim10^{8}$ km from Earth, would be ideal for such a new telescope to carry out these BH-stellar shadow observations.

As the telescope requirements for observing the shadow of a BH-stellar binary will not be fulfilled for at least several decades, we perform a secondary study of the light curves produced by these shadows. To observe the light curves, a VLBI is no longer required and many of the other requirements are relaxed, making such an observation more feasible in the near future. We calculate the light curves by integrating the photons from each image generated by the general relativistic ray-tracing code at many points along the binary's orbit. The number of photons is then normalized and plotted as a function of the position in the orbit, providing a light curve of the shadow. Such light curves could be extracted from observations of BH-stellar binary systems and could be used to determine the properties of the BH and the system as a whole.

The remainder of this paper presents the details pertaining to these results. Section~\ref{sec:Kerr} presents the Kerr BH and properties that are relevant to the BH shadow observation. Section~\ref{sec:shadow} discusses the BH shadow, ideally and as it may be seen in nature. Section~\ref{sec:algo} details the algorithm and setup we use to calculate the shadow of a BH-stellar system. Section~\ref{sec:res} presents the BH shadow images produced for four configurations of interest. Section~\ref{sec:feas} discusses the feasibility of observing a BH-stellar binary system shadow and estimates the observing capabilities required. Section~\ref{sec:curves} presents an alternative observation using the light curve of the system and the variations due to the shadow. Section~\ref{sec:conc} concludes by summarizing our results and discussing the implications. Throughout this paper, we use the following conventions: the metric signature is $(-,+,+,+)$; Latin letters in index lists stand for spacetime indices; we use geometric units with $G=1=c$ (e.g.~$1 M_\odot$ becomes 1.477 km by multiplying by $G/c^2$ or $4.93\times10^{-6}$ s by multiplying by $G/c^3$), except where otherwise noted.

\section{Kerr Black Hole}
\label{sec:Kerr}

The Kerr metric describes the exterior of a rotating (uncharged) BH. More technically, this metric is the most general solution to the Einstein equations in vacuum, assuming stationarity and axisymmetry. By the no-hair theorems of General Relativity~\cite{israel,1971PhRvL..26..331C,hawking-uniqueness,1975PhRvL..34..905R}, the Kerr metric depends only on the BH mass $M$ and its BH spin parameter $a\equiv|\vec{J}|/M$, where $|\vec{J}|$ is its spin angular momentum. In a particular coordinate system, labeled Boyer-Lindquist in honor of the scientists that discovered it~\cite{Boyer:1966qh}, the line element is given by
\begin{eqnarray}
ds^2=&-\left(1-\frac{2Mr}{\Sigma}\right)dt^2 - \frac{4Mra\sin^2\theta}{\Sigma}dtd\phi+\frac{\Sigma}{\Delta}dr^2 
\nonumber \\
&+\Sigma d\theta^2 + \left(r^2+a^2+\frac{2Ma^2r\sin^2\theta}{\Sigma}\right)d\phi^2,
\end{eqnarray}
where one typically defines
\begin{eqnarray}
\Sigma \equiv r^2+a^2\cos^2\theta,
\\
\Delta \equiv r^2-2Mr+a^2.
\end{eqnarray}
Of particular importance for this work are the event horizon and photon sphere of the Kerr spacetime, and thus, although this is basic relativity material, we will briefly summarize these concepts below and we will establish notation in the process (see e.g.~\cite{Misner:1973cw} for more details). 

\subsection{Event Horizon}

The event horizon is defined as a null surface created by marginally-trapped, null geodesics. Since the normal to the surface $n^{a}$ must be therefore null, the event horizon satisfies the horizon equation
\begin{equation}
g^{ab}\partial_{a}F\partial_{b}F=0,\label{eq:hor}
\end{equation}
where $g_{ab}$ is the spacetime metric and $F(x^{a})$ is a level surface function with normal $n_{a}=\partial_{a}F$. 

The Kerr spacetime is stationary, axisymmetric, and reflection symmetric about the poles and the equator, and thus the level surfaces are only dependent on radius. Without loss of generality, we let $F(x^{a})=r-r_{\hor}$, where $F=0$ defines the location of the horizon. Equation~\ref{eq:hor} becomes $g^{rr}=0$, and solving this equation one finds
\begin{equation}
r_{\hor}=M+(M^2-a^2)^{1/2}.
\end{equation}
The event horizon, thus, depends on the mass and the spin parameter, reducing to the familiar non-rotating (Schwarzschild) result $r_{\hor} = 2 M$ in the $a \to 0$ limit. 

\subsection{Trajectory of Photons}
\label{sec:ph-tra}

We start with the Hamilton-Jacobi equation
\begin{equation}
\frac{\partial S}{\partial\lambda} = \frac{1}{2}g^{ab}\frac{\partial S}{\partial x^{a}}\frac{\partial S}{\partial x^{b}}, \label{eq:HJ}
\end{equation}
where $S$ is the Jacobi action, $\lambda$ is the affine parameter, and $x^{a}$ are generalized coordinates. Using that the Kerr metric is stationary, axisymmetric and separable, the Jacobi action for null geodesics can be written as
\begin{equation}
S=-Et+L_{z}\phi+S_{r}(r)+S_{\theta}(\theta),
\end{equation}
where $E$ and $L_{z}$ are the specific energy and the $z$-component of the specific angular momentum, respectively. These are conserved quantities in the Kerr spacetime and are related to components of the four-momentum of a test particle in orbit in this spacetime via $p_{t}=-E$ and $p_{\phi}=L_{z}$. 

We insert the above ansatz into Eq.~\ref{eq:HJ}, 
\begin{eqnarray}
2\frac{\partial S}{\partial\lambda}=0=&g^{tt}E^{2}-2g^{t\phi}EL_{z}+g^{\phi\phi}L_{z}^{2}
\nonumber \\
&+g^{rr}\left(\frac{dS_{r}}{dr}\right)^{2}+g^{\theta\theta}\left(\frac{dS_{\theta}}{d\theta}\right)^{2},
\end{eqnarray}
and using separation of variables we find
\begin{eqnarray}
\Delta\left(\frac{dS_{r}}{dr}\right)^{2}=&\frac{1}{\Delta}[E(r^{2}+a^{2})+aL_{z}]^{2}-(L_{z}-aE)^{2}-\mathcal{Q},
\\
\left(\frac{dS_{\theta}}{d\theta}\right)^{2}=&\mathcal{Q}+\cos^{2}\theta\left[a^{2}E^{2}-L_{z}^{2}\csc^{2}\theta\right],
\end{eqnarray}
where $\mathcal{Q}$ is the so-called Carter constant. This quantity is a third conserved quantity of the Kerr spacetime, 
given in terms of the four-moment of a test particle via
\begin{equation}
\mathcal{Q}=p_{\theta}^{2}-\cos^{2}\theta(a^{2}E^{2}-L_{z}^{2}\csc^{2}\theta).
\end{equation}

Using now that $dS/dr=p_{r}=g_{rr}(dr/d\lambda)$ and $dS/d\theta=p_{\theta}=g_{\theta\theta}(d\theta/d\lambda)$,  the Hamilton-Jacobi equations become
\begin{eqnarray}
\Sigma\frac{dr}{d\lambda}=&\pm\sqrt{\mathcal{R}},\label{eq:dr}
\\
\Sigma\frac{d\theta}{d\lambda}=&\pm\sqrt{\Theta},\label{eq:dtheta}
\end{eqnarray}
where
\begin{eqnarray}
\mathcal{R}(r)\equiv&\left[E(r^{2}+a^{2})-aL_{z}\right]^{2}-\Delta\left[\mathcal{Q}+(L_{z}-aE)^{2}\right], \label{eq:R}
\\
\Theta(\theta)\equiv&\mathcal{Q}+\cos^{2}\theta[a^{2}E^{2}-L_{z}^{2}\csc^{2}\theta].
\end{eqnarray}
These are the null-geodesic equations for the $r(\lambda)$ and $\theta(\lambda)$ components of the trajectories of null particles.

\subsection{Photon Sphere}
\label{sec:ph-sph}

The photon sphere is defined as the the surface created by all unstable and spherical photon orbits, i.e.~the separatrix between photon geodesics that escape to spatial infinity and those that fall into the event horizon~\cite{Claudel:2000yi}. As the Kerr metric is separable, the photon sphere can be found analytically, as we do here.

Mathematically, the unstable spherical photon orbits are defined by the conditions $\mathcal{R}=0$, $d\mathcal{R}/dr=0$, and $\Theta\geq0$. Using the first two conditions and Eq.~\ref{eq:R} we can solve for the conserved quantities $\xi\equiv L_{z}/E$ and $\eta\equiv\mathcal{Q}/E^{2}$
\begin{eqnarray}
\xi_{\sph}=&\frac{r_{\sph}^{2}+a^2}{a}-\frac{2\Delta r_{\sph}}{a(r_{\sph}-M)}, \label{eq:xi}
\\
\eta_{\sph}=&-\frac{r_{\sph}^{3}\left[r_{\sph}(r_{\sph}-3M)^{2}-4a^{2}M\right]}{a^{2}(r_{\sph}-M)^{2}}, \label{eq:eta}
\end{eqnarray}
where $r_{\sph}$ is the constant radius of the unstable spherical photon orbits. 

This radius is constrained by the condition $\Theta \geq 0$, which can by rewritten in terms of $\xi$ and $\eta$ as
\begin{equation}
\frac{\Theta}{E^{2}}=\mathcal{J}-(a\sin\theta-\xi\csc\theta)^{2},
\end{equation}
where
\begin{equation}
\mathcal{J}\equiv\eta+(a-\xi)^{2}.
\end{equation}
Thus, we can see that there is a necessary (but not sufficient) condition for unstable spherical orbits of $\mathcal{J}\geq 0$. Substituting Eqs.~\ref{eq:xi} and~\ref{eq:eta} into the definition for $\mathcal{J}$ gives the condition
\begin{equation}
\mathcal{J}=\frac{4r_{\sph}^{2}\Delta}{(r_{\sph}-M)^{2}}\geq 0,
\end{equation}
which reduces to $\Delta\geq 0$ or $r_{\sph}\geq M+\sqrt{M^{2}-a^{2}}$, i.e.~the Kerr horizon radius. Clearly, there are no spherical photon orbits inside the event horizon.

We have found a set of equations and conditions that define all unstable spherical photon orbits. For each $r_{\sph}$ greater than the event horizon radius one can find $\xi_{\sph}$ and $\eta_{\sph}$ using Eqs.~\ref{eq:xi} and~\ref{eq:eta}, and in turn the null geodesic equations describing the motion of the photon, making sure to satisfy the condition on $\Theta$. All the unstable spherical photon orbits together form the photon sphere and the surface that divides the photons that will travel to spatial infinity and the photons that will fall into the event horizon.

As an example, we can solve for the radius of the unstable circular photon orbit that is in the equatorial plane. Since the orbit is confined to this plane, we have the additional conditions $\theta=\pi/2$ and $d\theta/d\lambda=0$. Thus, the Carter constant, as well as $\Theta$, are zero and the condition $\Theta\geq 0$ is satisfied. Then, using Eq.~\ref{eq:R} and the conditions $\mathcal{R}=0=d\mathcal{R}/dr$ we find
\begin{equation}
r_{\cir}=2M\left\{1+\cos\left[\frac{2}{3}\cos^{-1}\left(\mp\frac{a}{M}\right)\right]\right\},
\end{equation}
where the $(-)$ sign is for prograde orbits and the $(+)$ sign is for retrograde orbits. For a non-spinning BH, i.e.~the Schwarzschild spacetime, $r_{\cir}=3M$, as expected. For a maximally-spinning BH, i.e.~$a\rightarrow M$, $r_{\cir}\rightarrow M$ for prograde orbits and $r_{\cir}\rightarrow 4M$ for retrograde orbits.

\section{Black Hole Shadow}
\label{sec:shadow}

The BH shadow is an observable consequence of the BH photon sphere. Imagine an ideal situation where photons impinge on the BH isotropically from spatial infinity ($r=+\infty$). Any photons that cross the photon sphere will inexorably fall into the event horizon, and thus, it will never be seen by an outside observer. The remaining photons will follow a curved path near the BH, some getting deflected after the first passage, and some orbiting the BH a finite number of times and then returning to spatial infinity. An observer capturing the photons that escape the BH will therefore see a black region created by all the photons that crossed the photon sphere and fell into the event horizon: this is the BH shadow. Realistically, photons do not originate from spatial infinity, are not isotropically distributed, and do not return to spatial infinity unimpeded. We will discuss realistic BH shadows in later subsections.

\subsection{Ideal Shadows}

\begin{figure}
\includegraphics[width=\columnwidth,trim={7cm 7cm 7cm 7cm},clip]{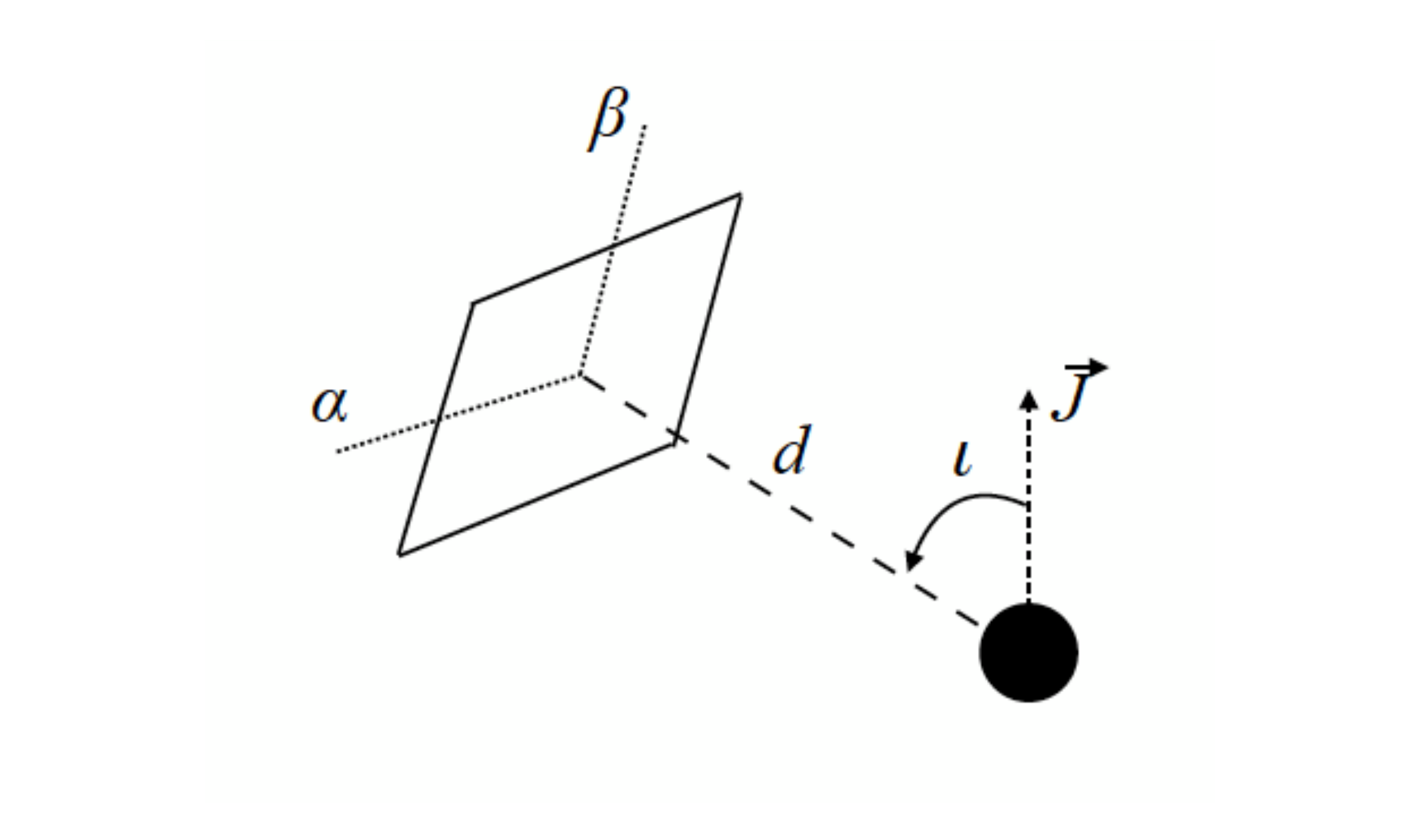}
\caption{\label{fig:obsscreen} Depiction of the orientation of the observing screen relative to the BH. $\alpha$ and $\beta$ are the celestial coordinates of the observing screen, $d$ is the distance to the BH, $\vec{J}$ is the spin angular momentum of the BH, and $\iota$ is the inclination angle between $\vec{J}$ and the observer's line of sight.}
\end{figure}

Since the ideal shadow is a direct consequence of the BH photon sphere it can be calculated analytically, just like the photon sphere is. Let us begin by defining a few angles that determine the location of the BH relative to the observer. We place the observer at spatial infinity with an orientation relative to the BH's spin angular momentum set by an inclination angle $\iota$, i.e.~the angle between the observer's line of sight and and the BH rotation axis. We then define the celestial coordinates ($\alpha$, $\beta$) as the apparent angular distances of the object on the celestial sphere as seen by the observer. These angles are measured along the directions perpendicular and parallel, respectively, to the rotation axis of the BH when projected onto the celestial sphere. Figure~\ref{fig:obsscreen} depicts the celestial coordinates on an observing screen relative to a BH with spin angular momentum $\vec{J}$. The celestial coordinates can be written in terms of the photon momentum as
\begin{eqnarray}
\alpha=&\lim_{r\rightarrow\infty}\frac{-rp^{(\phi)}}{p^{(t)}},
\\
\beta=&\lim_{r\rightarrow\infty}\frac{rp^{(\theta)}}{p^{(t)}},
\end{eqnarray}
where $p^{(a)}$ denotes the components of the photon four-momentum with respect to a locally non-rotating reference frame~\cite{1972ApJ...178..347B}. $p^{(a)}$ is related to the general photon four-momentum $p^{a}$ by a coordinate transformation (e.g.~$p^{\phi}=p^{(\phi)}/\sin\iota$). The celestial coordinates can also be written in terms of the conserved parameters $\xi$ and $\eta$:
\begin{eqnarray}
\alpha=&-\frac{\xi_{\sph}}{\sin\iota},
\\
\beta=&\left(\eta_{\sph}+a^{2}\cos^{2}\iota-\xi_{\sph}^{2}\cot^{2}\iota\right)^{1/2}.
\end{eqnarray}

\begin{figure}[hpt]
\includegraphics[width=\columnwidth,clip]{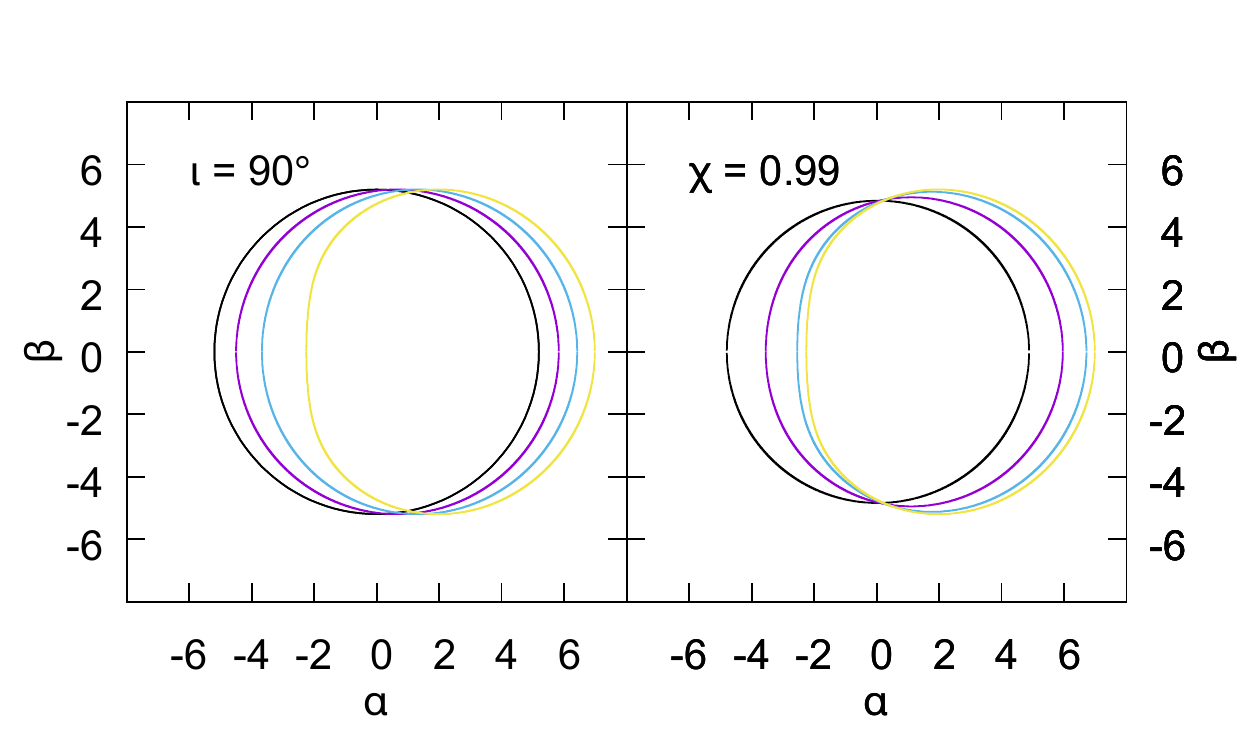}
\caption{\label{fig:kerrshadow}(Color Online) Boundary of the Kerr black hole shadow for various values of dimensionless spin $\chi=a/M$ and inclination angle $\iota$. The left figure fixes $\iota=90\degree$, and, from left to right, varies the spin through $\chi=0$ (black), $0.33$ (purple), $0.66$ (blue), $0.99$ (yellow). The right plot fixes $\chi=0.99$, and, from left to right, varies the inclination angle through $\iota=0\degree$ (black), $30\degree$ (purple), $60\degree$ (blue), $90\degree$ (yellow).}
\end{figure}

With this at hand, we can now compute the BH shadows as a function of the celestial coordinates. As we did in the calculation of the photon sphere, we vary $r_{\sph}$ to calculate the innermost photon orbit, i.e.~the outer boundary of the shadow, in terms of the celestial coordinates and as seen by the observer. Figure~\ref{fig:kerrshadow} shows the BH shadow for various values of the spin parameter and the inclination angle.
Observe that as the spin is increased for fixed inclination angle (or as the inclination angle is increased for fixed spin), the shadow is deformed ands shifted to the right. Therefore, it stands to reason that if one were to measure the BH shadow, one might be able to extract the spin and inclination angle of the BH observed. 

\subsection{Accretion Disk Shadows}

In reality, the primary source of photons impinging on a BH will be some nearby luminous object. The most commonly considered source is a BH accretion disk, i.e.~a disk of gas and dust orbiting the BH~\cite{lrr-2013-1}. Supermassive BHs, the primary candidates for current BH shadow observations~\cite{2009astro2010S..68D, 2008SPIE.7013E..2AE, Goddi:2017pfy}, are found at the centers of galaxies, and thus, they have a large source of dust and gas to form an accretion disk. For stellar mass BHs, the disk formation process is a bit more complicated. Since the majority of massive stars are found in binary systems~\cite{2008ASPC..387...93G}, it is not unreasonable for a stellar mass BH to be found in a binary with a massive star. If the orbital separation is small enough, the gravitational pull of the BH will transfer gas from the star to an orbit around the BH, forming a disk over time. For larger orbital separations and more massive stars, it is possible the stellar wind of the star will transfer a significant amount of gas to the BH, also forming an accretion disk~\cite{2017RvMP...89b5001B}.

The material in accretion disks moves on approximately Keplerian orbits around the BH, with some small inward radial motion (small as compared with the angular velocity). Particles that are at smaller radii have larger angular velocities, i.e.~the disk is \emph{differentially rotating}. This differential rotation leads to viscous torque between particles at different radii, which transfers angular momentum outwards and heats up the inner radii of the disk~\cite{2002apa..book.....F}. The hot disk thus radiates thermally, and along with contributions from Compton scattering, synchrotron emission (if the disk has a significant magnetic field)~\cite{1973blho.conf..343N}, and a reflection component (if there is a hot corona)~\cite{2017RvMP...89b5001B}, the disk can be a fairly luminous object that sources the photons for a BH shadow observation.

Unlike the ideal case discussed earlier, photons from an accretion disk do not originate from spatial infinity and are not isotropically distributed. Therefore, these BH shadows are influenced by the physics of the disk. Moreover, since the disk is very near the BH it will itself contribute to any BH shadow image, for example obscuring part of the shadow. Such effects on the shadow have been studied~\cite{Psaltis:2014mca, Chan:2014nsa, Medeiros:2016put, 2017ApJ...850..172J}, and will need to be taken into account when observations are analyzed.

\subsection{Binary Shadows}

A much less commonly considered source of photons for a BH shadow is a companion star in a binary configuration with the BH. As mentioned previously, the majority of massive stars are found in binary systems and so it is not unreasonable to consider a low-mass stellar-BH binary. In general, though, an accretion disk will dominate as a source of photons over the companion star, primarily because it is much closer to the BH. But if an accretion disk has not formed (or is sparse), because the orbital separation is still relatively large, the star can be the primary source of illumination for the BH shadow. These kinds of systems are difficult to identify and currently there is only one candidate system, 2M05215658+4359220~\cite{Thompson:2018ycv}, but as it was only recently discovered more work is required to confirm its nature.

The shadow produced with a star as the primary source would be significantly different than both the ideal shadow and the shadow sourced by an accretion disk. Photons from the star certainly do not approach the BH isotropically and the direction they do approach from changes with time, since the BH and star are in orbit about their center of mass, which is close to the BH singularity if the BH is much more massive than the star. The resultant shadow will be time-dependent and likely very asymmetric. It will also depend on the properties of the binary system, such as the orbital eccentricity, the semi-major axis, and the relative angles between the orbital plane, the BH rotation axis, and the observer's line of sight. All of these features of a binary BH shadow make it an interesting target for study and we do so in the following sections.

\section{Numerical Methodology}
\label{sec:algo}

In this section, we describe our numerical algorithm for calculating the BH shadow produced in a low-mass stellar-BH binary system. To reduce the computational cost, we split our method in two parts. The first is a general relativistic ray-tracing code that computes the trajectories of photons near the BH. We initialize the photon trajectories directed at the BH from an observing screen far from it. We then numerically solve the geodesic equations to find and store all the photon trajectories, in particular storing the direction (relative to the BH) that photons had to originate from if they were to pass near the BH and continue on to hit the observing screen. The second part of the method selects which of these photons would have originated from a star in a binary orbit around the BH. The first part of the method only depends on the spin of the BH and the inclination angle, while the second depends on the physical parameters of the binary system, such as the orbital separation, eccentricity, and the radius of the star. By splitting the numerical method in this way, we can reuse the photon trajectory data with a given BH spin and inclination angle for a variety of stellar and binary parameters.

Our general relativistic ray-tracing code follows the method laid out in~\cite{Psaltis:2010ww}. As discussed in Sec.~\ref{sec:ph-sph}, in the Kerr spacetime (as well as in any stationary and axisymmetric spacetime) the orbital energy $E$ and $z$-component of the orbital angular momentum $L_{z}$ are related to components of the photon four-momentum by $p_{t}=-E$ and $p_{\phi}=L_{z}$. These relations can be rewritten to give us two first-order differential equations for the $t$ and $\phi$ components of the photon four-position
\begin{eqnarray}
\frac{dt}{d\lambda'}=&-\frac{\xi g_{t\phi}+g_{\phi\phi}}{g_{tt}g_{\phi\phi}-g_{t\phi}^{2}},\label{eq:dt}
\\
\frac{d\phi}{d\lambda'}=&\xi\frac{g_{tt}+g_{t\phi}}{g_{tt}g_{\phi\phi}-g_{t\phi}^{2}},\label{eq:dphi}
\end{eqnarray}
where $\lambda'\equiv E\lambda$ is the normalized affine parameter and $\xi\equiv L_{z}/E$ is the same conserved quantity as in Sec.~\ref{sec:ph-sph}.

For the $r$ and $\theta$ components of the photon position we could use the first-order differential equations given in Eqs.~\ref{eq:dr} and~\ref{eq:dtheta}, but to keep our ray-tracing code applicable to spacetimes that may not be separable we instead use the second-order geodesic equations for a general axisymmetric metric:
\begin{eqnarray}
\frac{d^{2}r}{d\lambda'^{2}}=&-\Gamma^{r}_{tt}\left(\frac{dt}{d\lambda'}\right)^{2}-\Gamma^{r}_{rr}\left(\frac{dr}{d\lambda'}\right)^{2}-\Gamma^{r}_{\theta\theta}\left(\frac{d\theta}{d\lambda'}\right)^{2}
\nonumber \\
&-\Gamma^{r}_{\phi\phi}\left(\frac{d\phi}{d\lambda'}\right)^{2}-2\Gamma^{r}_{t\phi}\left(\frac{dt}{d\lambda'}\right)\left(\frac{d\phi}{d\lambda'}\right)
\nonumber \\
&-2\Gamma^{r}_{r\theta}\left(\frac{dr}{d\lambda'}\right)\left(\frac{d\theta}{d\lambda'}\right),\label{eq:d2r}
\\ \nonumber \\
\frac{d^{2}\theta}{d\lambda'^{2}}=&-\Gamma^{\theta}_{tt}\left(\frac{dt}{d\lambda'}\right)^{2}-\Gamma^{\theta}_{rr}\left(\frac{dr}{d\lambda'}\right)^{2}-\Gamma^{\theta}_{\theta\theta}\left(\frac{d\theta}{d\lambda'}\right)^{2}
\nonumber \\
&-\Gamma^{\theta}_{\phi\phi}\left(\frac{d\phi}{d\lambda'}\right)^{2}-2\Gamma^{\theta}_{t\phi}\left(\frac{dt}{d\lambda'}\right)\left(\frac{d\phi}{d\lambda'}\right)
\nonumber \\
&-2\Gamma^{\theta}_{r\theta}\left(\frac{dr}{d\lambda'}\right)\left(\frac{d\theta}{d\lambda'}\right),\label{eq:d2theta}
\end{eqnarray}
where $\Gamma^{a}_{bc}$ are the Christoffel symbols of the metric.

We use a reference frame and coordinate system such that the BH is stationary at the origin and the BH's spin angular momentum is along the $z$-axis. For the numerical evolution, the observing screen is centered at a distance $d=1000$, where we use ``code units'' with the BH mass $M=1$, the azimuthal angle $\theta=\iota$, and the polar angle $\phi=0$. On the screen, we use the polar coordinates $r_{\scr}$ and $\phi_{\scr}$, which are related to the observer's celestial coordinates by $\alpha=r_{\scr}\cos\phi_{\scr}$ and $\beta=r_{\scr}\sin\phi_{\scr}$. The orientation of the screen and BH is depicted in Fig.~\ref{fig:obsscreen}. 

We evolve our system of equations backwards in time since we know the final positions and momenta of the photons that hit the screen, but do not know where they originated from. Thus, we initialize each photon at some point on the screen and with a four-momentum that is perpendicular to the screen. The latter simulates the observing screen being at spatial infinity since only those photons that are moving perpendicular to the screen at a distance $d$ would also hit the screen at spatial infinity. 

With this at hand, the initial position and four-momentum of each photon in the coordinates of the BH spacetime is given by
\begin{eqnarray}
r_{i}=&\left(d^{2}+\alpha^{2}+\beta^{2}\right)^{1/2},
\\
\theta_{i}=&\arccos\left(\frac{d\cos\iota+\beta\sin\iota}{r_{i}}\right),
\\
\phi_{i}=&\arctan\left(\frac{\alpha}{d\sin\iota-\beta\cos\iota}\right),
\end{eqnarray}
and
\begin{eqnarray}
\left(\frac{dr}{d\lambda'}\right)_{i}&=\frac{d}{r_{i}},
\\
\left(\frac{d\theta}{d\lambda'}\right)_{i}&=\frac{-\cos\iota+\frac{d}{r_{i}^{2}}\left(d\cos\iota+\beta\sin\iota\right)}{\sqrt{r_{i}^{2}-\left(d\cos\iota+\beta\sin\iota\right)^{2}}},
\\
\left(\frac{d\phi}{d\lambda'}\right)_{i}&=\frac{-\alpha\sin\iota}{\alpha^{2}+\left(d\sin\iota-\beta\cos\iota\right)^{2}},
\\
\left(\frac{dt}{d\lambda'}\right)_{i}&=-\frac{g_{t\phi}}{g_{tt}}\left(\frac{d\phi}{d\lambda'}\right)_{i}-\left[\frac{g_{t\phi}^{2}}{g_{tt}^{2}}\left(\frac{d\phi}{d\lambda'}\right)_{i}^{2}\right.
\nonumber \\
&\left.-\left(g_{rr}\left(\frac{dr}{d\lambda'}\right)_{i}^{2}+g_{\theta\theta}\left(\frac{d\theta}{d\lambda'}\right)_{i}^{2}+g_{\phi\phi}\left(\frac{d\phi}{d\lambda'}\right)_{i}^{2}\right)\right]^{1/2},
\end{eqnarray}
where to find $\left(dt/d\lambda'\right)_{i}$ we used that the norm of the photon four-momentum is zero.  Completing the initialization, we compute the conserved quantity $\xi$ from the initial conditions, as this quantity is required in Eqs.~\ref{eq:dt} and~\ref{eq:dphi}.

We vary the initial positions and momenta of the photons by varying $r_{\scr}$ over the range $0\leq r_{\scr}\leq10$ in steps of $10^{-6}$ and $\phi_{\scr}$ over the range $0\leq\phi_{\scr}\leq2\pi$ in steps of $\pi/180$. The small step-size on the screen ensures that we have an ample sampling of the possible photon paths. A Runge-Kutta-Fehlberg method is used to numerically evolve the system of differential equations given by Eqs.~\ref{eq:dt},~\ref{eq:dphi},~\ref{eq:d2r}, and~\ref{eq:d2theta}. If the photon crosses $r=r_{\hor}+\delta r$, where we use $\delta r=10^{-5}$, we stop the evolution and do not record the final position of the photon. In this case, any photon that hit that location on the screen would have had to originate from inside the event horizon which is not possible. If the photon evolves to a radius $r>d$, we stop the evolution and record the final position of the photon. This position gives us the direction from which the photon would have originated, since the curvature due to the BH spacetime at $r>d$ is negligible, and thus any photons at this distance would be traveling radially outwards from the BH. The code is run separately for each set of BH spin and inclination angle parameters, generating a data set of initial and final photon positions for each set of parameters.

\begin{figure*}[hpt]
  \centering
    \includegraphics[width=0.4\columnwidth,trim={0 3cm 0 3cm},clip]{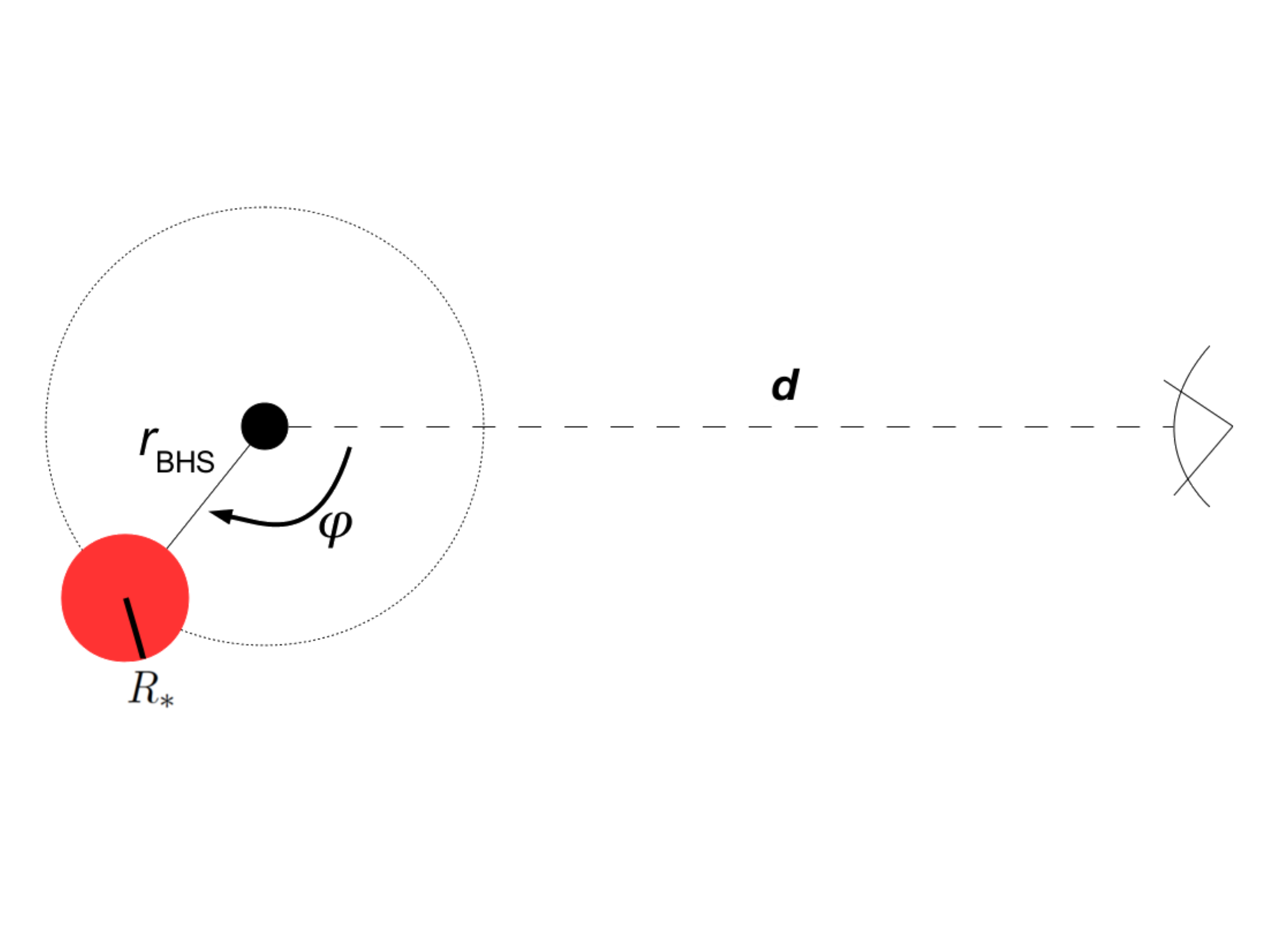}
    \includegraphics[width=0.4\columnwidth,trim={0, 3cm, 0, 3cm},clip]{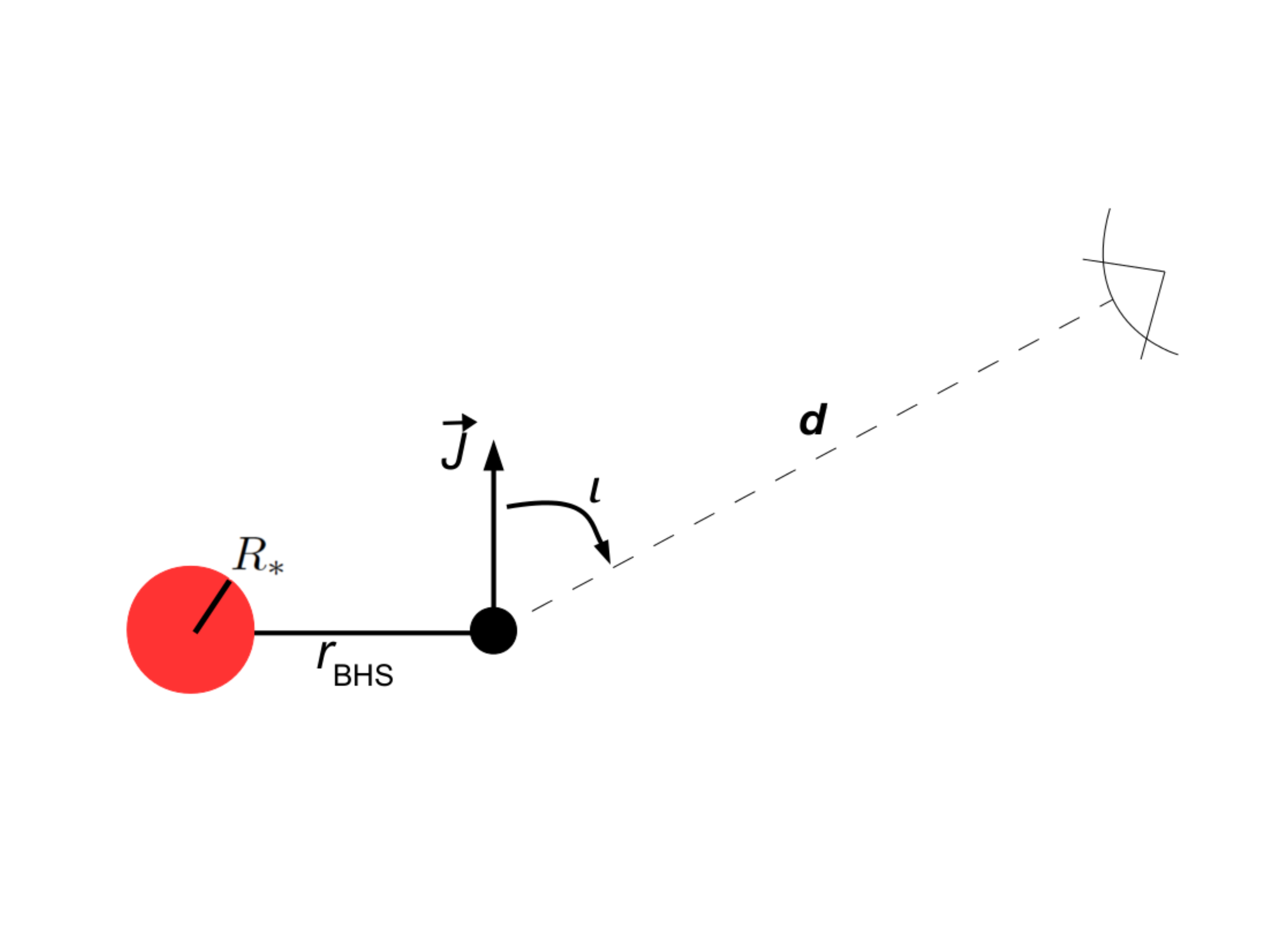}
    \caption{\label{fig:topview} Top-down (left) and edge-on (right) view of the stellar-BH binary system for a circular orbit (not to scale). The smaller, filled black circle represents the BH and the larger, filled red circle represents the star. The observer at distance $d$ from the system is also included with a line of sight directed at the BH at an inclination angle $\iota$ relative to the BH's spin angular momentum $\vec{J}$. The semi-major axis of the orbit is $r_{\BHS}$, while $\varphi$ is the angular location of the star in its orbit with $\varphi=0$ along the observer's line of sight when projected onto the orbital plane.}
\end{figure*}

The second part of our algorithm makes use of the physical parameters of the binary system and the star. The setup of the stellar-BH binary system is described in Fig.~\ref{fig:topview}. As with the ray-tracing code we use a reference frame and coordinate system such that the BH is stationary at the origin and the BH's spin angular momentum is along the $z$-axis. For simplicity we align the orbital angular momentum of the star with the BH's spin angular momentum\footnote{While the two angular momenta need not necessarily be exactly aligned, there is good evidence from numerical simulations that the majority of stellar-BH binary systems have small misalignment ($< 10\degree$)~\cite{2010ApJ...719L..79F}.}, as shown in the right-panel of Fig.~\ref{fig:topview}, but the code is written such that any alignment can be modeled. In Fig.~\ref{fig:topview} we depict a circular orbit, but we also model non-circular orbits. The semi-major axis of the orbit is denoted by $r_{\BHS}$ and $\varphi$ is the angular location of the star in its orbit, with $\varphi=0$ along the observer's line of sight when projected onto the orbital plane. We also place the periapsis at $\varphi=0$, but our code does not require this by default.

We model the orbit through a Keplerian ellipse with an orbital separation of $r_{\OS}$, given by
\begin{equation}
r_{\OS}=\frac{r_{\BHS}(1-e^{2})}{1+e\cos\varphi},\label{eq:orbsep}
\end{equation}
where $e$ is the orbital eccentricity. From $r_{\OS}$ and the radius of the star $R_{*}$ we compute the angular radius of the star $\vartheta$ as seen on the BH's sky
\begin{equation}
\vartheta=\arctan\left(\frac{R_{*}}{r_{\OS}}\right).\label{eq:angrad}
\end{equation}
Such a model for the binary orbit should suffice, since we assume that the orbital separation is large enough that general relativistic effects are ignorable. If we had chosen a small separation, such that relativistic effects were important, then a disk would form and an accretion disk shadow would be created instead of a binary shadow. 

Using the position of the star and its angular size we now determine which photons from the ray-tracing code have a final position that lies within the extent of the star on the sky of the BH. These photons are the ones that would have originated from the star and would have gone on to impact the observing screen at spatial infinity. Finally, we plot the observing screen positions of the photons that could have originated from the star to produce the binary BH shadow. To roughly approximate what an actual telescope may observe we divide the observing screen into an equal 40x40 grid and create a 2D histogram of the photon counts in each bin. Since the image of the BH shadow depends on the star's orbital angular position $\varphi$ we can in principle produce movies of how the shadow changes as the star orbits around the BH.

\section{Numerical Results}
\label{sec:res}

We present the results from four BH-star binary configurations here, although the BH shadows of configurations spanning the entire parameter space were produced as part of testing our algorithm. The parameters of the four configurations are summarized in Table~\ref{tab:configs}. Configuration 1 is a high spin configuration, Configuration 2 is a high eccentricity configuration, Configuration 3 is similar to the A0620-00 system~\cite{1986ApJ...308..110M, 2010ApJ...710.1127C}, and Configuration 4 is similar to the Cygnus X-1 system~\cite{Ziolkowski:2005ag, 2011ApJ...742...84O, Parker:2015fja}. Note that we show the angular radius of the star $\vartheta$ as a parameter rather than the radius of the star $R_{*}$ and the orbital semi-major axis $r_{\BHS}$. This is because only the ratio between $R_{*}$ and $r_{\BHS}$ is relevant for the shadow as can be seen from Eqs.~\ref{eq:orbsep} and~\ref{eq:angrad}. For non-circular configurations we show the star's angular radius as seen at apoapsis, i.e.~$\varphi=180\degree$. 

\begin{table}[hpt]
\begin{center}
\begin{tabular}{c || c | c | c | c}
Configuration & $\chi$ & $\iota$ & $e$ & $\vartheta$ \\ \hline
1 & $0.95$ & $90\degree$ & $0.0$ & $3.945\degree$ \\ \hline
2 &  $0.0$ & $90\degree$ & $0.9$ & $5.711\degree$ \\ \hline
3 &  $0.12$ & $57\degree$ & $0.0$ & $5.711\degree$ \\ \hline
4 & $0.95$ & $46\degree$ & $0.0018$ & $20.39\degree$ \\
\end{tabular}
\caption{\label{tab:configs} Table summarizing the parameters of the four configurations presented in this work. $\chi=a/M$ is the dimensionless spin parameter of the BH, $\iota$ is the inclination angle, $e$ is the orbital eccentricity, and $\vartheta$ is the angular radius of the star as seen at apoapsis (for comparison, the angular radius of the Sun and the Moon as seen from Earth are both roughly $0.25\degree$).}
\end{center}
\end{table}

\begin{figure*}
\includegraphics[width=0.5\columnwidth,trim={0 1cm 0 0},clip]{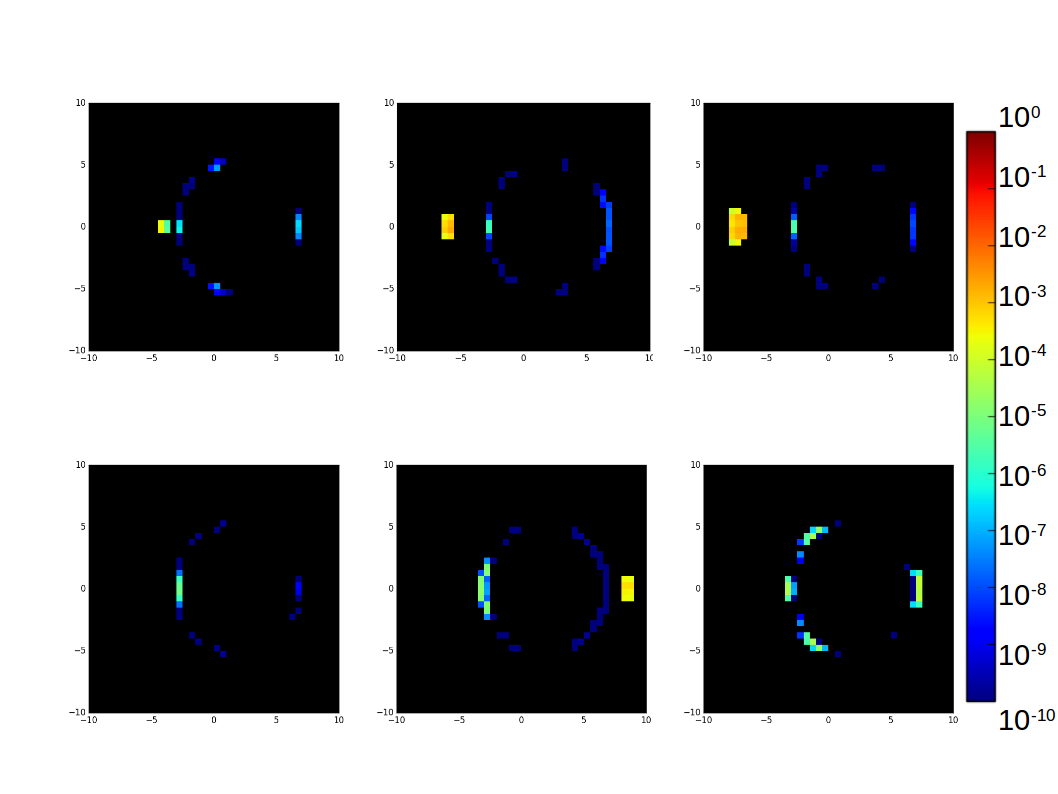}
\includegraphics[width=0.5\columnwidth,trim={0 1cm 0 0},clip]{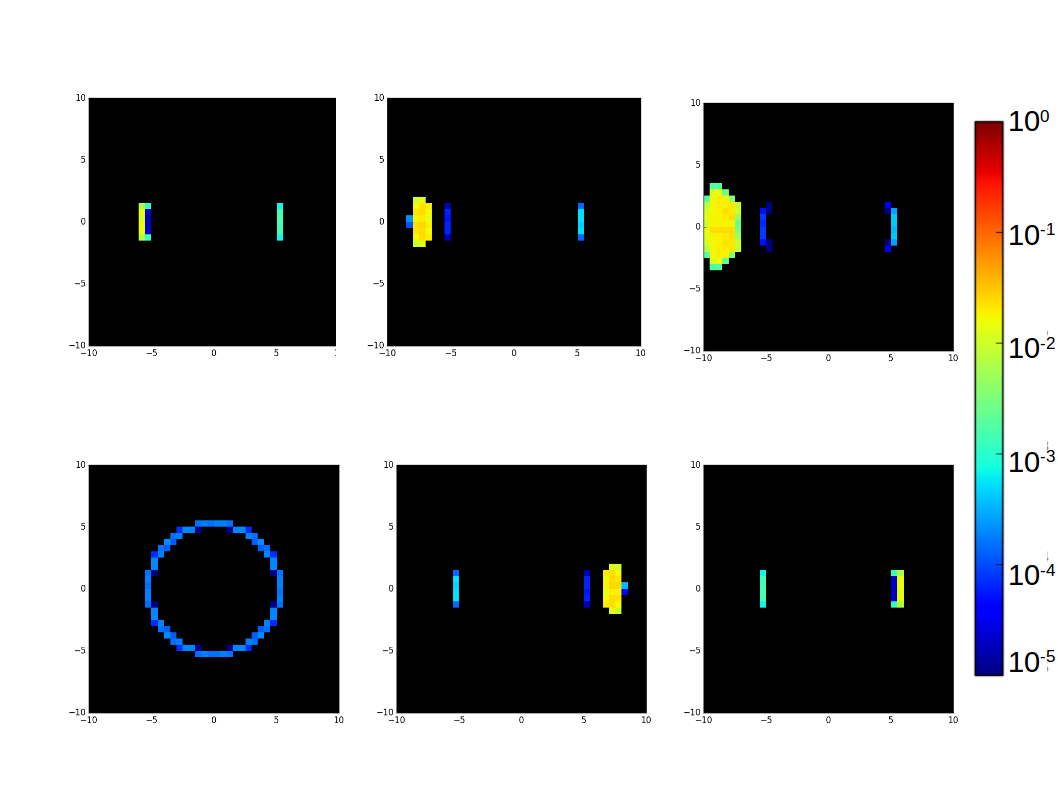}
\\
\includegraphics[width=0.5\columnwidth,trim={0 1cm 0 0},clip]{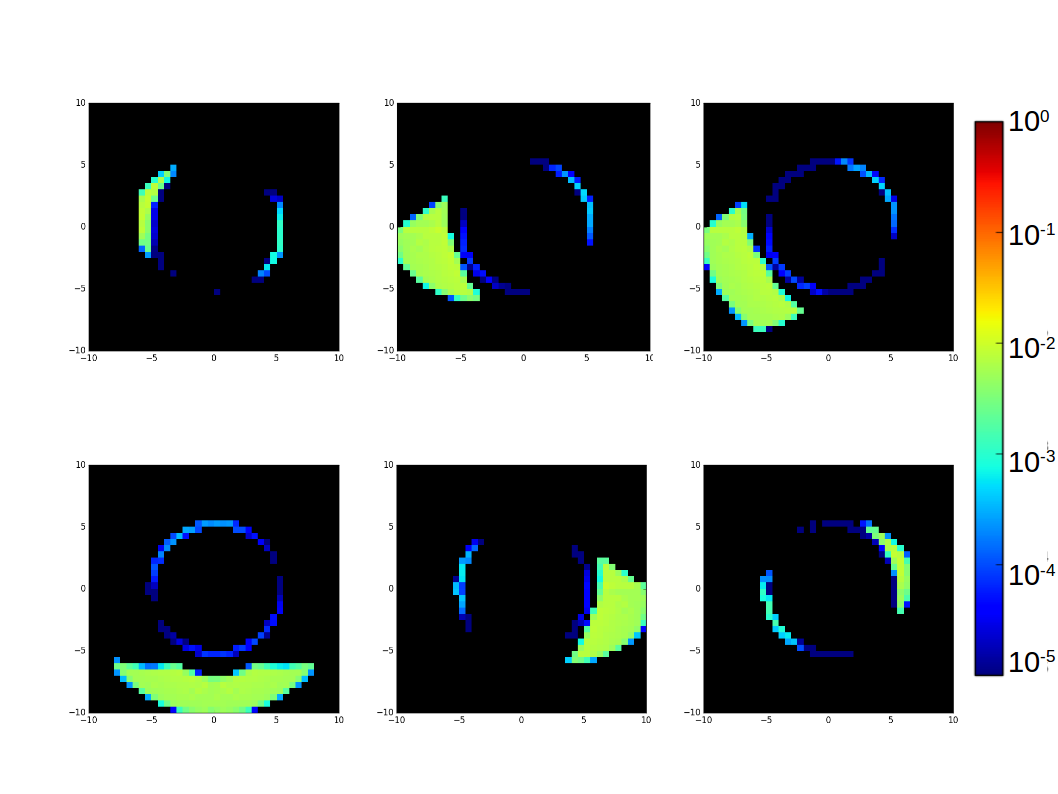}
\includegraphics[width=0.5\columnwidth,trim={0 1cm 0 0},clip]{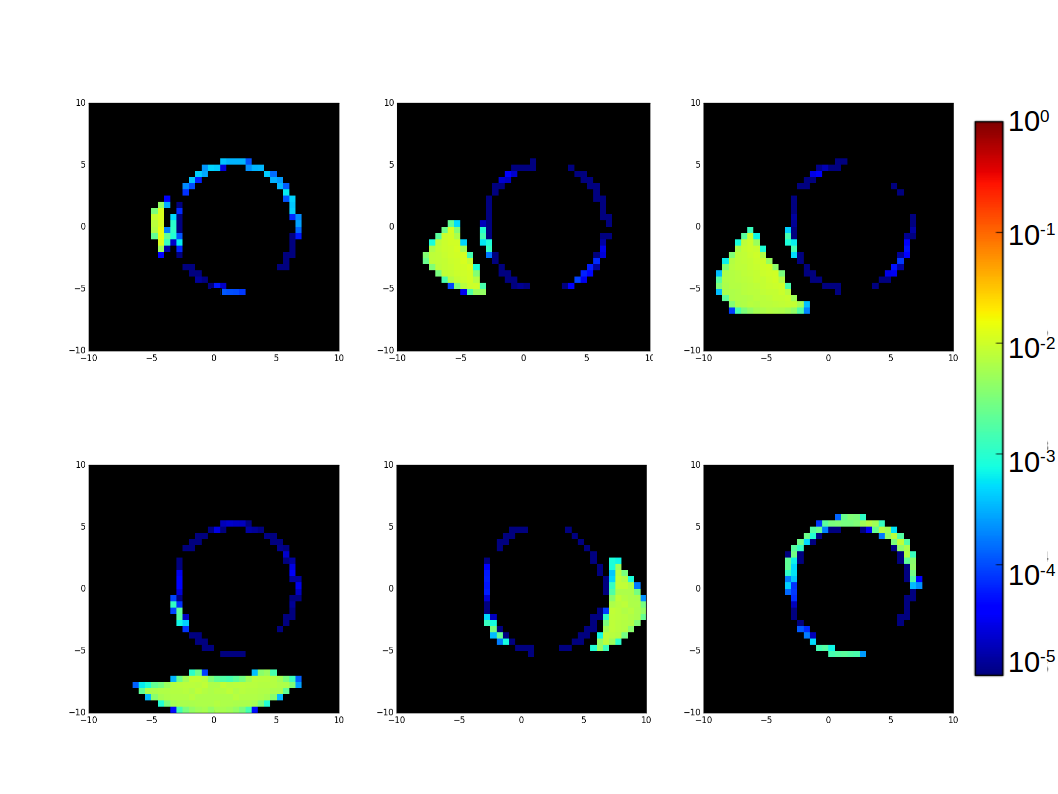}
\caption{\label{fig:shadows}BH shadow images for Configuration 1 (top left), 2 (top right) 3 (bottom left), and 4 (bottom right) with parameters described in Table~\ref{tab:configs}. For each configuration starting from top left the images are at $\varphi=60\degree$, $120\degree$, $135\degree$, $180\degree$, $240\degree$, and $300\degree$.  The color scale represents the ratio of photons in the bin vs the greatest number of photons in the star's orbit.}
\end{figure*}

As this is the first work studying BH shadows of BH-star binary systems, we will discuss our results qualitatively and leave a more detailed quantitative analysis for future work. Figure~\ref{fig:shadows} shows the BH shadows of the four configurations at $\varphi=60\degree$, $120\degree$, $135\degree$, $180\degree$, $240\degree$, and $300\degree$. Observe the complexity that is present in the images, especially considering that there is a different shadow for each location in the star's orbit. In principle, since these images contain a wealth of information about the system, the physical parameters of the BH, the star, and the binary system as a whole, may be extractable if the images can be observed with sufficient resolution.

As an example of possible parameter extraction, note that the extent of the images in Configurations 3 and 4 are significantly larger than the extent of the images in Configurations 1 and 2. This must be due, at least partially, to the larger angular size of the star in the BH's sky, which is in turn a function of the radius of the star and the orbital semi-major axis. Another feature of the images is their asymmetry, e.g.~in Configuration 1 at $\varphi=180\degree$ the image is not symmetric between the right and left half of the image. This asymmetry can only be due to the spin of the BH as no other parameters in our setup can lead to such an effect at this position in the orbit. A second type of asymmetry is that seen between images at opposing points in the orbit, i.e.~the pairs $\varphi=60\degree$ and $300\degree$ and $\varphi=120\degree$ and $240\degree$. For a circular orbit and an inclination angle of $\iota=0\degree$, i.e.~looking down the orbital and rotation axis, these pairs of images would be the same due to symmetry. But as seen in the configurations we are presenting, the spin of the BH, the orbital eccentricity, and the non-zero inclination angle, create an asymmetry in the images at opposing points in the orbit.

The ability to extract physical parameters would depend on the systematic error present in the modeling of such observations and the observational error present in the data from telescopes. The former would primarily be caused by the complexities of accurately modeling electromagnetic emission from a star and the assumption that an accretion disk is not present or does not contribute significantly to the BH shadow images. The observational error includes the statistical, instrumental, and environmental errors that would be present in any telescopes attempting to make such observations. How much these sources of error will limit the ability to extract parameters from BH shadow observations of BH-star binaries is left for future work, but in the following section we discuss whether these observations would be feasible in the future and what kind of observing capabilities would be required.

\section{Observational Feasibility}
\label{sec:feas}

In this section we estimate the feasibility of detecting and resolving the shadows of a BH-star binary system. We also discuss some of the major technological hurdles that need to be overcome before such an observing campaign will be possible. 

\subsection{Physical Requirements}

There are two dominant factors, that determine if it is possible to observe the shadows of a BH-star binary system. The first is whether the shadow is bright enough to be observable from Earth with current and/or future telescopes. This first factor primarily depends on the distance of the system from Earth, the mass of the BH, the luminosity and size of the companion star, and the orbital separation of the system. Resolving the BH shadow requires the use of VLBI techniques and to resolve objects with smaller angular sizes requires larger baselines, i.e.~distances between the telescopes in the VLBI setup. Thus, our second factor is if the baseline required to resolve angular sizes comparable to the BH shadow's angular size is feasible now or in the future. 

Let us then first consider the feasibility of detecting the shadow by estimating its magnitude. The bolometric apparent magnitude of any object is given by
\begin{equation}
m=-0.17+\log_{10}\left[\left(\frac{L}{L_{\odot}}\right)^{-2.5}\left(\frac{D}{1\text{ pc}}\right)^{5}\right],
\end{equation}
where $L_{\odot}$ is the Sun's luminosity, $L$ is the luminosity of the object, and $D$ is the distance to the object from Earth. We calculate the luminosity of the BH shadow by treating the BH as an object that isotropically scatters any light that enters some volume around the BH. We model this volume as a sphere of radius $r_{\hor}=2M$, i.e.~the BH event horizon radius for a non-rotating BH. The fraction of light emitted by the star that enters this volume, and thus the luminosity of the BH shadow, is given by
\begin{equation}
\frac{L_{\BH}}{L_{*}}=\frac{M^{2}}{r_{\OS}^{2}},
\end{equation}
where $L_{*}$ is the luminosity of the star. As we want to detect the shadow, its apparent magnitude must be smaller than some minimum detectable threshold $m^{*}$, which then leads to the criterion  
\begin{equation}
\log_{10}\left[\left(\frac{L_{*}}{L_{\odot}}\right)^{-2.5}\left(\frac{GM}{r_{\OS}c^{2}}\right)^{-5}\left(\frac{D}{1\text{ pc}}\right)^{5}\right] \lesssim m^{*} + 0.17,
\label{eq:BHmag}
\end{equation}
where we have restored factors of $G$ and $c$ in this equation. Note that this calculation gives the bolometric apparent magnitude of the whole shadow, and thus, it is a very optimistic estimate of the required observing capabilities for spatially resolving the shadow. In reality, the total luminosity of the shadow will be divided over the whole image, and so any sub-region of the shadow will have a lower luminosity and a larger apparent magnitude, requiring a more sensitive telescope.

Let us now estimate the feasibility of resolving the BH shadow, which clearly depends on its angular size and the VLBI baseline. The angular radius of the shadow as seen from Earth can be estimated from the radius of the photon sphere for a Schwarzschild BH, i.e.~$r_{\cir}=3M$. The angular radius is then
\begin{equation}
\vartheta_{\BH}=\arctan\left(\frac{3M}{D}\right).\label{eq:BHangrad}
\end{equation}
The angular resolution can be approximated by
\begin{equation}
\Delta\vartheta=\frac{\lambda}{d},
\end{equation}
where $\lambda$ is the wavelength of observed light and $d$ is the diameter of the telescope's objective, i.e.~the element that gathers and focuses the light from the observed object. For a telescope array as used in VLBI, the diameter of the telescope's objective is replaced with the array's largest baseline $B$.  As we want to resolve features of the BH shadow, we need an angular resolution that is smaller than the angular size of the shadow. In Fig.~\ref{fig:shadows}, the binning used gives an angular resolution that is roughly $1/10$ the radius of the shadow boundary, i.e.~$\Delta\vartheta\approx\vartheta_{\BH}/10$, and so to remain consistent with that calculation, we use this resolution here as well. The required VLBI baseline is then given by
\begin{equation}
B \gtrsim \frac{\lambda}{\arctan\left(\frac{3GM}{10Dc^{2}}\right)},\label{eq:base}
\end{equation}
where again we have restored factors of $G$ and $c$.

Let us now evaluate these criteria assuming one wished to observe the BH shadows of our known, example BH-star binary systems. Table~\ref{tab:detect} summarizes the relevant properties of three of the nearest BH-star binary systems and the only non-interacting BH-star binary candidate\footnote{Some of the properties for these systems are not well constrained and/or there is disagreement in their values. We use values that are roughly centered in the accepted range.}, as well as the angular radius of the shadow, the apparent magnitude of the shadow, and the baseline required to resolve the shadow, calculated from Eqs.~\ref{eq:BHangrad},~\ref{eq:BHmag}, and~\ref{eq:base}, respectively. We set $\lambda = 550$ nm for the observing wavelength $\lambda$, a value that is in the middle of the optical range. We thereby make the simplifying assumption that a filter centered in the optical would capture the majority of the black hole shadow's bolometric luminosity. Note that we include the three nearest BH-star binary systems as representative of non-interacting systems that may be found in the future, but do not suggest they are candidates for such an observation as they contain an accretion disk.
\begin{center}
\begin{table}[hpt]
\begin{tabular}{c || c | c | c | c}
System & $M (M_{\odot})$ & $D(\text{kpc})$ & $r_{\OS}(\text{AU})$ & $L_{*}(L_{\odot})$ \\ \hline
A 0620-00\cite{1986ApJ...308..110M, 2010ApJ...710.1127C} & $6.6$ & $1.1$ & $0.013$ & $0.3$ \\ \hline
Cygnus X-1\cite{Ziolkowski:2005ag, 2011ApJ...742...84O, Parker:2015fja} &  $15$ & $1.8$ & $0.2$ & $3.5\times10^{5}$ \\ \hline
V404 Cygni\cite{1994MNRAS.271L..10S, 1994MNRAS.271L...5C, Bernardini:2016mub} & $9$ & $2.4$ & $0.16$ & $10.2$ \\ \hline
2M05215658+4359220\cite{Thompson:2018ycv} & $4$ & $3.7$ & $0.71$ & $400$ \\
\end{tabular}
\begin{tabular}{c || c | c | c}
System & $\vartheta_{\BH}(\text{$\mu$as})$ & $m$ & $B(10^{6}\text{ km})$ \\ \hline
A 0620-00\cite{1986ApJ...308..110M, 2010ApJ...710.1127C} & $1.8\times10^{-4}$ & $43$ & $6.4$ \\ \hline
Cygnus X-1\cite{Ziolkowski:2005ag, 2011ApJ...742...84O, Parker:2015fja} & $2.5\times10^{-4}$ & $33$ & $4.6$ \\ \hline
V404 Cygni\cite{1994MNRAS.271L..10S, 1994MNRAS.271L...5C, Bernardini:2016mub} & $1.3\times10^{-4}$ & $45$ & $10.0$ \\ \hline
2M05215658+4359220\cite{Thompson:2018ycv} & $3.2\times10^{-5}$ & $47$ & $35.4$ \\
\end{tabular}
\caption{\label{tab:detect} Properties of three of the nearest BH-star binary systems. $M$ is the BH mass, $D$ is the distance to the system, $r_{\OS}$ is the orbital separation at apoapsis, and $L_{*}$ is the luminosity of the companion star. The angular radius of the shadow $\vartheta_{\BH}$ is calculated from Eq.~\ref{eq:BHangrad}. The apparent magnitude of the shadow $m$ is calculated from Eq.~\ref{eq:BHmag}. The baseline required to resolve the shadow $B$ is calculated from Eq.~\ref{eq:base}.}
\end{table}
\end{center}

Let us first discuss the apparent magnitude of the BH shadows of these three systems. Cygnus X-1 is the brightest by far with an apparent magnitude of $m=33$, which is expected given that its companion star is a blue supergiant with a significantly higher luminosity than the K-type stars of A 0620-00 and V404 Cygni. Currently, the faintest objects have been observed with the Hubble Space Telescope (HST) in the eXtreme Deep Field (XDF) data set, which combines data from 10 years of observations for a total observing time of about 50 days~\cite{2013ApJS..209....6I}. These objects have an apparent magnitude of $m\approx31$. NASA is considering a successor for HST to launch in the 2030s called the Large UV/Optical/Infrared Surveyor (LUVOIR)~\cite{LUVOIR}. If we only consider the improvement in telescope aperture, LUVOIR will be 10-40 times more sensitive than Hubble, which corresponds to the ability to observe objects with apparent magnitudes larger by 2-4. Given the observing capabilities available to us today and in the near future, it is thus not unreasonable to assume that in the next few decades it will be possible to construct and launch space telescopes that can observe BH shadows as dim as that of Cygnus X-1's.

Let us now discuss the baseline requirement. The Event Horizon Telescope~\cite{2009astro2010S..68D} is currently the largest VLBI telescope with a baseline on the order of $B\approx10^{4}$ km giving it an angular resolution on the order of $\Delta\vartheta\approx1$ $\mu$as (EHT observes at a wavelength of $\lambda=1.3$ mm). EHT is about the limit of what is capable for purely Earth-based telescopes, so to resolve the shadows of BH-star binary systems requires a space-based VLBI telescope. Radioastron~\cite{Kardashev:2013bca} is the first and currently only space-based VLBI with a baseline about an order of magnitude larger than EHT. This baseline is still two orders of magnitude too small to resolve the shadows of BH-star binaries, but it does show that a space-based VLBI is possible. In the future, a more advanced space-based VLBI could be set up between a space telescope at one of the Lagrange points of the Earth-Sun system (there are already several space telescopes at the two nearest Lagrange points, $L_{1}$ and $L_{2}$, and more are planned for the future, such as LUVOIR). $L_{1}$ and $L_{2}$ are too close to the Earth to provide a large enough baseline, but $L_{4}$ and $L_{5}$, which are $60\degree$ ahead or behind Earth's orbit, are about $1.5\times10^{8}$ km from Earth. A telescope placed at $L_{4}$ or $L_{5}$ would form a baseline with telescopes on Earth that is more than large enough to resolve the BH shadows of the three BHs we have considered here.

\subsection{Technological Requirements}
\label{sec:tech}

In the previous section, we set the observing wavelength to the middle of the optical range, $\lambda=550$ nm. This was done for two reasons: the emission from the star peaks in the optical (and thus, so does the shadow image) and the required VLBI baseline is proportional to the observing wavelength. If instead one were to use a wavelength in the radio range, such as $\lambda=1$ mm as used by EHT, the baseline would need to be over $3$ orders of magnitude larger, comparable to the size of the solar system. More importantly, if we approximate the star as a blackbody, the luminosity in the radio range vanishes very quickly. Developing a space-based interferometer that can be placed at such far distances from Earth and that can detect such low luminosities is not feasible in the near or far future. For this reason, we choose the observing wavelength to be in the optical range, but doing so brings along other technological problems that are not necessarily surmountable at present in radio interferometers, but which could be surmountable in the future. We discuss the more prominent ones below.

In radio VLBI with very large baselines, such as that in EHT, the electromagnetic signal observed by each telescope must be digitized and recorded to later be transmitted/transported to and processed at a central location. This is currently not possible for optical VLBI because the required sampling rates are beyond modern technology. Thus, optical VLBI is limited to baselines of several hundred kilometers so that the original observed signals can be transmitted through fiber cables to a central location to be combined and processed directly on site. With these current limitations in mind, we can make a rough approximation of the sampling rates that would be required for optical VLBI. The EHT observes at a frequency of 230 GHz and has sampling rates of 8 giga-samples/sec~\cite{1538-3873-130-986-044502}, which are currently the highest rates of any VLBI telescope. Assuming the required rate is proportional to the observing frequency and leaving all other factors equivalent, we can estimate that an optical VLBI would require a rate on the order of 10 tera-samples/sec, 3 orders of magnitude higher than what is available with current technology. These sampling rates, over about the past decade, have been doubling roughly every 2 years. Assuming that rate of advancement continues, it is possible that sampling rates on the order of 10 tera-samples/sec will be feasible in 2 to 3 decades.

Along with a high sampling rate comes the significant amount of data that is produced and needs to be transmitted to a central location to be processed. For VLBI telescopes with large baselines on Earth, this is usually done by recording the data onto hard drives and physically transporting the hard drives to the central location. Obviously, this is not possible for a space-based VLBI telescope and the data would instead have to be transmitted using the communications systems of each space telescope. For simplicity, let us assume the telescopes must have transmission rates that match their data recording rates. With a sampling rate of 10 tera-samples/sec and 2 bits per sample (as in EHT~\cite{1538-3873-130-986-044502}), the data recording rate and the required transmission rate is 20 terabits/sec. The Deep Space Network, the primary communications system for NASA missions, currently has a maximum transmission rate of about 5 megabits/sec~\cite{DSN}. NASA expects and is planning for future missions to require transmission rates that increase by about an order of magnitude each decade, but has already tested new technologies allowing for transmission rates approaching 1 gigabits/sec. Given that outlook, it is not unreasonable to expect transmission rates on the order of 10 terabits/sec within the next century, if not in the next 50 years.

There is, in principle, an alternative to overcoming the technological requirements discussed above. What has been discussed so far applies for a type of interferometry called \textit{amplitude} and/or \textit{phase interferometry}, in which two (or more) electromagnetic signals are combined and the resulting interference pattern can be used to extract information about the original signals. Another type of interferometry, called \textit{intensity interferometry}, cross-correlates quantum fluctuations in the signal between different telescopes, with the amount of correlation, or ``de-correlation'', providing information on the original signal~\cite{2016SPIE.9907E..0MD}. Intensity interferometry has the benefit of requiring significantly lower sampling rates as compared to phase interferometry, but it has the drawback that to create an image from observations requires a large array of telescopes. A space-based version of such an interferometer would require an array of space telescopes spread over a large area, and estimating when such an array would be possible is difficult. In addition, while an intensity interferometer was successfully tested in the 1970s, the technique has been essentially non-existent in astronomy since then~\cite{2016SPIE.9907E..0MD}. In the past few years, there has been a resurgence of interest in intensity interferometry, particularly with the ongoing construction of the Cherenkov Telescope Array~\cite{Ong:2017ihp}, which is well suited for this imaging technique. Intensity interferometry is of course still effectively in its infancy and much more must be done before space-based arrays are even considered, but it does provide another avenue to reaching the imaging capabilities required for observing the shadows of BH-star binary systems.

\section{Light Curve of the Shadow}
\label{sec:curves}

As discussed, there are a number of difficulties associated with observing the BH shadow of a BH-star binary system. One can relax the requirement for a VLBI, and in turn the technological requirements discussed in Sec.~\ref{sec:tech}, with an alternative observation that can still provide a wealth of information about the system. Rather than spatially resolving the shadow and studying the images, we can also study the simple photometry of the entire shadow. Over the course of the binary orbit, this creates a light curve with characteristic variations. These variations will depend on the variations in the star itself, and on the effect of the BH shadow. In principle, the light curve variations due to the shadow are the only ones that will be periodic, with a period equivalent to the orbital period, and thus, they could be separated from the remaining variation. Detecting the shadow variations would still require a telescope large enough to observe apparent magnitudes about equal to that of the shadow itself, but the observation can be done with a single Earth- or space-based telescope rather than a space-based VLBI array. In this section, we calculate and discuss the light curve variations associated with the BH shadows presented in Sec.~\ref{sec:res}.

\begin{figure*}
\includegraphics[width=0.5\columnwidth,trim={0 1cm 0 0},clip]{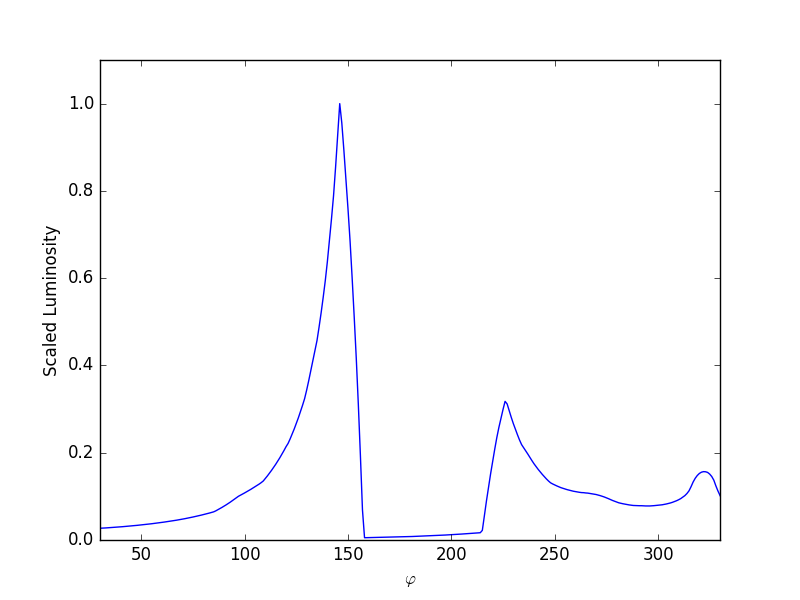}
\includegraphics[width=0.5\columnwidth,trim={0 1cm 0 0},clip]{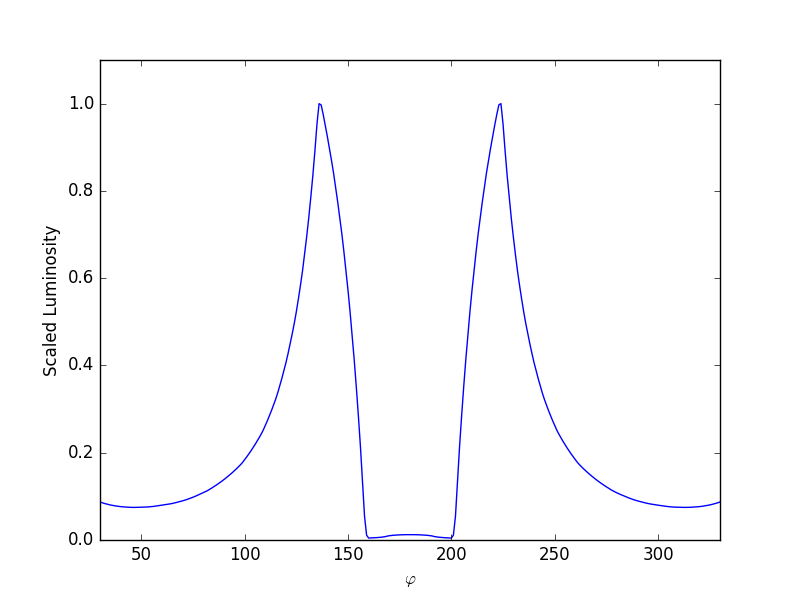}
\\
\includegraphics[width=0.5\columnwidth,trim={0 1cm 0 0},clip]{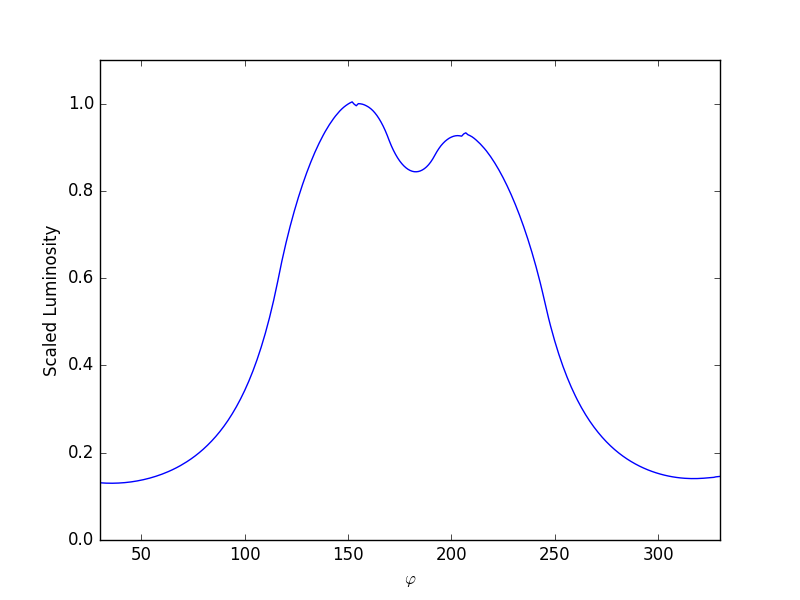}
\includegraphics[width=0.5\columnwidth,trim={0 1cm 0 0},clip]{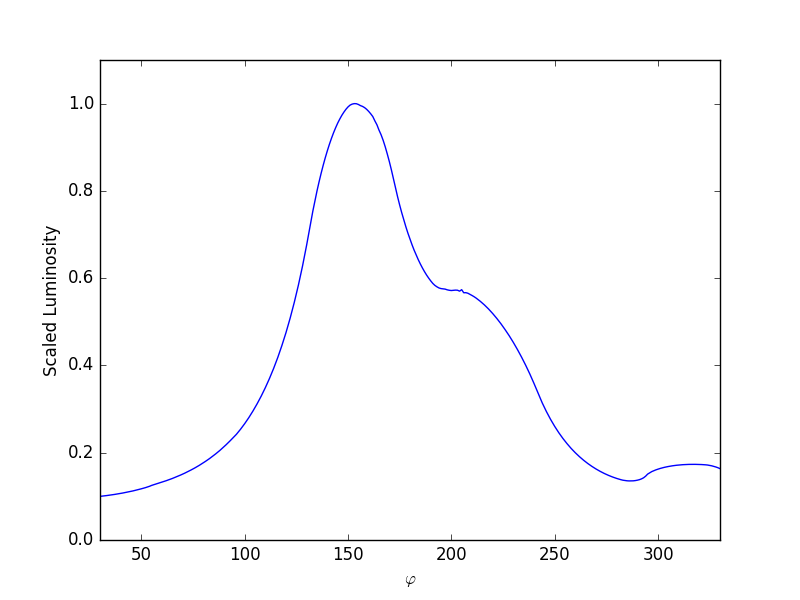}
\caption{\label{fig:curves}BH shadow light curves for Configuration 1 (top left), 2 (top right) 3 (bottom left), and 4 (bottom right) with parameters described in Table~\ref{tab:configs}. The light curves are scaled such that the brightest point of the light curve is equal to 1.}
\end{figure*}

We calculate the light curves, and in turn the light curve variations, of the BH shadow for each configuration simply by totaling the number of photons in each image at orbit angles $30\degree\leq\varphi\leq330\degree$. The remaining angles are excluded as the star begins to enter the viewing screen and obscures the shadow. This effect would need to be taken into account when analyzing a real observation, but we exclude it as we wish to focus on the shadow itself. We then scale the light curves such that the largest photon count is equal to 1, and plot them as a function of the orbit angle. Figure~\ref{fig:curves} shows the BH shadow light curves for the BH-star binary configurations described in Sec.~\ref{sec:res}. One can at first glance notice that the light curve for each configuration is quite distinct, suggesting that it may be possible to extract a fair amount of information about the system and the BH.

There are some features evident in the light curves that can be attributed to the physical parameters of the system and the BH. The effect of the inclination angle can be clearly seen when comparing Configurations 1 and 2 with Configurations 3 and 4. In Configurations 1 and 2 the inclination angle is $\iota=90\degree$,~i.e.~the system is being viewed edge-on. Thus, at an orbital angle of $\varphi=180\degree$,~i.e.~when the star is directly behind the BH along the line of sight, only the photons that pass very near to the BH are visible and so the luminosity is relatively very low. In the case of Configurations 3 and 4, however, the systems are viewed at some angle and so the star is not occluded by the BH at $\varphi=180\degree$. The secondary larger and brighter lensed image of the star is still visible, as can be seen in Fig.~\ref{fig:shadows}, and the luminosity of the shadow remains relatively high. The spin of the BH also plays a role as can be seen in Configurations 1, 3, and 4. The light curves in these cases are asymmetric, especially in Configurations 1 and 4 where the spin is very high $\chi=0.95$. Even in Configuration 3 which has a low spin $\chi=0.12$, the first peak in the light curve is slightly larger than the second. This asymmetry must be due to the BH spin as there are no other physical parameters in our simulation setup that can cause the asymmetry. In reality, asymmetry could be caused by the eccentricity and other orbital parameters, but we leave such studies to future work. 

As with the BH shadow images, the ability to extract physical parameters depends not only on statistical covariances between the parameters in the model, but also on the systematic error in the modeling and the observational error in the data. While the light curves are a simpler observation than the shadow, there will still be systematic error introduced by the difficulty of accurately modeling electromagnetic emission from a star and by the assumption that this is a non-interacting BH-star binary with no accretion disk. The observational error for shadow light curves is of particular importance as the light curve of the shadow will appear as a very small variation on the light curve of the system. Any large sources of observational error can overshadow or distort the shadow light curves and make it very difficult or impossible to perform any parameter extraction. We leave a more detailed analysis of the ability to extract parameters using the light curves of BH shadows of BH-star binaries for future work.

\section{Conclusions and Future Directions}
\label{sec:conc}

We have performed the first study of BH shadows produced by BH-star binary systems, where the companion star is the primary source of electromagnetic radiation. We produced images of what the BH shadows would look like to some future VLBI telescope with four configurations, one of which is similar to the Cygnus X-1 system and another to the A0620-00 system. We argue that due to the complexity of the shadows and their predictable time-dependence, their observations could be used to estimate the parameters of the BH, the star, and the system as a whole. We estimated the apparent magnitude and angular size of three of the nearest BH-star binary systems~\cite{1986ApJ...308..110M, 2010ApJ...710.1127C, Ziolkowski:2005ag, 2011ApJ...742...84O, Parker:2015fja, 1994MNRAS.271L..10S, 1994MNRAS.271L...5C, Bernardini:2016mub} and the only non-interacting BH-star binary candidate~\cite{Thompson:2018ycv} and compared those quantities to the capabilities of current and upcoming telescopes. We argued that given current and upcoming capabilities it is not unreasonable to think that observing the BH shadows of BH-star binary systems will be possible within the next century, if not sooner. We also calculated the light curves of BH shadows as a more feasible alternative to imaging and resolving the shadows themselves. We found that, even though light curves are a simpler observation, they could still be used to estimate parameters of the BH, the star, and the system.

Since to our knowledge this is the first study of its kind, our analysis of BH shadows has been qualitative, but much more work could be done to explore their scientific case more thoroughly for a wider range of configurations. In particular, our claim that observations of these shadows could be used to estimate parameters of the system could be studied more carefully with an actual data analysis investigation. Perhaps shadow observations of such systems would be well suited for tests of GR, seeing as the shadows of supermassive BHs are already a prime target for such studies~\cite{Psaltis:2014mca, Vincent:2016sjq, 2017RvMP...89b5001B}.

In producing our shadow images we made a number of simplifying assumptions that could also be relaxed in future work. One such simplification is our treatment of the star as a source of electromagnetic radiation. Through our algorithm the star is effectively a two-dimensional, monochromatic, and isotropically emitting source. A more realistic model of emission could be implemented as well as a ray-tracing algorithm that actually traces rays to the star's surface. The simplification of using monochromatic light also extends to the final images we produced and the calculations for determining the feasibility of these shadow observations. In our calculations, we work with the bolometric luminosity, but stars do not emit at a single wavelength and telescopes have a limited range of observable wavelengths. Extending the ray-tracing code and feasibility calculations to include the effects of non-monochromatic light and non-ideal telescopes would provide more realistic images and capability requirements for observing the shadows of BH-star binary systems.

\ack

NY, DA, and HG acknowledge support from the NSF CAREER grant PHY-1250636. NY and HG also acknowledge support from NASA grants NNX16AB98G and 80NSSC17M0041. DA also acknowledges support from the National Natural Science Foundation of China (NSFC), Grant No. U1531117, and Fudan University, Grant No. IDH1512060. We would also like to acknowledge the support of the Research Group at Montana State University through their High Performance Computer Cluster \emph{Hyalite}.  

\section*{References}
\bibliography{biblio}

\providecommand{\newblock}{}
\begin{thebibliography}{10}
\expandafter\ifx\csname url\endcsname\relax
  \def\url#1{{\tt #1}}\fi
\expandafter\ifx\csname urlprefix\endcsname\relax\def\urlprefix{URL }\fi
\providecommand{\eprint}[2][]{\url{#2}}

\bibitem{2009astro2010S..68D}
{Doeleman} S, {Agol} E, {Backer} D, {Baganoff} F, {Bower} G~C, {Broderick} A,
  {Fabian} A, {Fish} V, {Gammie} C, {Ho} P, {Honman} M, {Krichbaum} T, {Loeb}
  A, {Marrone} D, {Reid} M, {Rogers} A, {Shapiro} I, {Strittmatter} P,
  {Tilanus} R, {Weintroub} J, {Whitney} A, {Wright} M and {Ziurys} L 2009
  {Imaging an Event Horizon: submm-VLBI of a Super Massive Black Hole} {\em
  astro2010: The Astronomy and Astrophysics Decadal Survey\/} ({\em
  Astronomy\/} vol 2010) (\textit{Preprint} \eprint{0906.3899})

\bibitem{Goddi:2017pfy}
Goddi C {\em et~al.\/} 2016 {\em Int. J. Mod. Phys.\/} {\bf D26} 1730001
  [1,863(2017)] (\textit{Preprint} \eprint{1606.08879})

\bibitem{2008SPIE.7013E..2AE}
{Eisenhauer} F, {Perrin} G, {Brandner} W, {Straubmeier} C, {Richichi} A,
  {Gillessen} S, {Berger} J~P, {Hippler} S, {Eckart} A, {Sch{\"o}ller} M,
  {Rabien} S, {Cassaing} F, {Lenzen} R, {Thiel} M, {Cl{\'e}net} Y, {Ramos} J~R,
  {Kellner} S, {F{\'e}dou} P, {Baumeister} H, {Hofmann} R, {Gendron} E, {Boehm}
  A, {Bartko} H, {Haubois} X, {Klein} R, {Dodds-Eden} K, {Houairi} K, {Hormuth}
  F, {Gr{\"a}ter} A, {Jocou} L, {Naranjo} V, {Genzel} R, {Kervella} P,
  {Henning} T, {Hamaus} N, {Lacour} S, {Neumann} U, {Haug} M, {Malbet} F,
  {Laun} W, {Kolmeder} J, {Paumard} T, {Rohloff} R~R, {Pfuhl} O, {Perraut} K,
  {Ziegleder} J, {Rouan} D and {Rousset} G 2008 {GRAVITY: getting to the event
  horizon of Sgr A*} {\em Optical and Infrared Interferometry\/} ({\em
  \procspie\/} vol 7013) p 70132A (\textit{Preprint} \eprint{0808.0063})

\bibitem{1973ApJ...183..237C}
{Cunningham} C~T and {Bardeen} J~M 1973 {\em \apj\/} {\bf 183} 237--264

\bibitem{Dokuchaev:2018fze}
Dokuchaev V~I and Nazarova N~O 2017 {\em JETP Lett.\/} {\bf 106} 637--642
  [Pisma Zh. Eksp. Teor. Fiz.106,no.10,609(2017)] (\textit{Preprint}
  \eprint{1802.00817})

\bibitem{1986ApJ...308..110M}
{McClintock} J~E and {Remillard} R~A 1986 {\em \apj\/} {\bf 308} 110--122

\bibitem{2010ApJ...710.1127C}
{Cantrell} A~G, {Bailyn} C~D, {Orosz} J~A, {McClintock} J~E, {Remillard} R~A,
  {Froning} C~S, {Neilsen} J, {Gelino} D~M and {Gou} L 2010 {\em \apj\/} {\bf
  710} 1127--1141 (\textit{Preprint} \eprint{1001.0261})

\bibitem{Ziolkowski:2005ag}
Ziolkowski J 2005 {\em Mon. Not. Roy. Astron. Soc.\/} {\bf 358} 851--859
  (\textit{Preprint} \eprint{astro-ph/0501102})

\bibitem{2011ApJ...742...84O}
{Orosz} J~A, {McClintock} J~E, {Aufdenberg} J~P, {Remillard} R~A, {Reid} M~J,
  {Narayan} R and {Gou} L 2011 {\em \apj\/} {\bf 742} 84 (\textit{Preprint}
  \eprint{1106.3689})

\bibitem{Parker:2015fja}
Parker M~L {\em et~al.\/} 2015 {\em Astrophys. J.\/} {\bf 808} 9
  (\textit{Preprint} \eprint{1506.00007})

\bibitem{1994MNRAS.271L..10S}
{Shahbaz} T, {Ringwald} F~A, {Bunn} J~C, {Naylor} T, {Charles} P~A and
  {Casares} J 1994 {\em \mnras\/} {\bf 271} L10--L14

\bibitem{1994MNRAS.271L...5C}
{Casares} J and {Charles} P~A 1994 {\em \mnras\/} {\bf 271} L5--L9

\bibitem{Bernardini:2016mub}
Bernardini F, Russell D~M, Shaw A~W, Lewis F, Charles P~A, Koljonen K~I~I,
  Lasota J~P and Casares J 2016 {\em Astrophys. J.\/} {\bf 818} L5
  (\textit{Preprint} \eprint{1601.04550})

\bibitem{Thompson:2018ycv}
Thompson T~A {\em et~al.\/} 2018  (\textit{Preprint} \eprint{1806.02751})

\bibitem{LUVOIR}
Rioux N, Thronson H~A, Feinberg L~D, Stahl H~P, Redding D~C, Jones A~L, Sturm
  J~A, Collins C~M, Liu A~K~C and Bolcar M~R 2016 {\em Journal of Astronomical
  Telescopes, Instruments, and Systems\/} {\bf 2} 2 -- 2 -- 9
  \urlprefix\url{https://doi.org/10.1117/1.JATIS.2.4.041214}

\bibitem{israel}
Israel W 1967 {\em Phys. Rev.\/} {\bf 164} 1776--1779

\bibitem{1971PhRvL..26..331C}
{Carter} B 1971 {\em Physical Review Letters\/} {\bf 26} 331--333

\bibitem{hawking-uniqueness}
Hawking S~W 1972 {\em Commun. Math. Phys.\/} {\bf 25} 152--166

\bibitem{1975PhRvL..34..905R}
{Robinson} D~C 1975 {\em Physical Review Letters\/} {\bf 34} 905

\bibitem{Boyer:1966qh}
Boyer R~H and Lindquist R~W 1967 {\em J. Math. Phys.\/} {\bf 8} 265

\bibitem{Misner:1973cw}
Misner C~W, Thorne K~S and Wheeler J~A 1973 {\em Gravitation\/} (San Francisco:
  W.H. Freeman)

\bibitem{Claudel:2000yi}
Claudel C~M, Virbhadra K~S and Ellis G~F~R 2001 {\em J. Math. Phys.\/} {\bf 42}
  818--838 (\textit{Preprint} \eprint{gr-qc/0005050})

\bibitem{1972ApJ...178..347B}
{Bardeen} J~M, {Press} W~H and {Teukolsky} S~A 1972 {\em \apj\/} {\bf 178}
  347--370

\bibitem{lrr-2013-1}
Abramowicz M~A and Fragile P~C 2013 {\em Living Reviews in Relativity\/} {\bf
  16} \urlprefix\url{http://www.livingreviews.org/lrr-2013-1}

\bibitem{2008ASPC..387...93G}
{Gies} D~R 2008 {Binaries in Massive Star Formation} {\em Massive Star
  Formation: Observations Confront Theory\/} ({\em Astronomical Society of the
  Pacific Conference Series\/} vol 387) ed {Beuther} H, {Linz} H and {Henning}
  T p~93

\bibitem{2017RvMP...89b5001B}
{Bambi} C 2017 {\em Reviews of Modern Physics\/} {\bf 89} 025001
  (\textit{Preprint} \eprint{1509.03884})

\bibitem{2002apa..book.....F}
{Frank} J, {King} A and {Raine} D~J 2002 {\em {Accretion Power in Astrophysics:
  Third Edition}\/}

\bibitem{1973blho.conf..343N}
{Novikov} I~D and {Thorne} K~S 1973 {Astrophysics of black holes.} {\em Black
  Holes\/} ed {Dewitt} C and {Dewitt} B~S pp 343--450

\bibitem{Psaltis:2014mca}
Psaltis D, Ozel F, Chan C~K and Marrone D~P 2015 {\em Astrophys. J.\/} {\bf
  814} 115 (\textit{Preprint} \eprint{1411.1454})

\bibitem{Chan:2014nsa}
Chan C~K, Psaltis D, Ozel F, Narayan R and Sadowski A 2015 {\em Astrophys.
  J.\/} {\bf 799} 1 (\textit{Preprint} \eprint{1410.3492})

\bibitem{Medeiros:2016put}
Medeiros L, Chan C~k, Ozel F, Psaltis D, Kim J, Marrone D and Sadowski A 2017
  {\em Astrophys. J.\/} {\bf 844} 35 (\textit{Preprint} \eprint{1610.03505})

\bibitem{2017ApJ...850..172J}
{Johnson} M~D, {Bouman} K~L, {Blackburn} L, {Chael} A~A, {Rosen} J, {Shiokawa}
  H, {Roelofs} F, {Akiyama} K, {Fish} V~L and {Doeleman} S~S 2017 {\em \apj\/}
  {\bf 850} 172 (\textit{Preprint} \eprint{1711.01286})

\bibitem{Psaltis:2010ww}
Psaltis D and Johannsen T 2012 {\em Astrophys. J.\/} {\bf 745} 1
  (\textit{Preprint} \eprint{1011.4078})

\bibitem{2010ApJ...719L..79F}
{Fragos} T, {Tremmel} M, {Rantsiou} E and {Belczynski} K 2010 {\em \apjl\/}
  {\bf 719} L79--L83 (\textit{Preprint} \eprint{1001.1107})

\bibitem{2013ApJS..209....6I}
{Illingworth} G~D, {Magee} D, {Oesch} P~A, {Bouwens} R~J, {Labb{\'e}} I,
  {Stiavelli} M, {van Dokkum} P~G, {Franx} M, {Trenti} M, {Carollo} C~M and
  {Gonzalez} V 2013 {\em \apjs\/} {\bf 209} 6 (\textit{Preprint}
  \eprint{1305.1931})

\bibitem{Kardashev:2013bca}
Kardashev N~S, Kovalev Y~Y and Kellermann K~I 2012 {\em URSI Radio Science
  Bulletin\/} {\bf 343} 22--29 (\textit{Preprint} \eprint{1303.5200})

\bibitem{1538-3873-130-986-044502}
MacMahon D~H~E, Price D~C, Lebofsky M, Siemion A~P~V, Croft S, DeBoer D,
  Enriquez J~E, Gajjar V, Hellbourg G, Isaacson H, Werthimer D, Abdurashidova
  Z, Bloss M, Brandt J, Creager R, Ford J, Lynch R~S, Maddalena R~J, McCullough
  R, Ray J, Whitehead M and Woody D 2018 {\em Publications of the Astronomical
  Society of the Pacific\/} {\bf 130} 044502
  \urlprefix\url{http://stacks.iop.org/1538-3873/130/i=986/a=044502}

\bibitem{DSN}
JPL N Deep space communications
  \urlprefix\url{https://scienceandtechnology.jpl.nasa.gov/research/research-topics-list/communications-computing-software/deep-space-communications}

\bibitem{2016SPIE.9907E..0MD}
{Dravins} D 2016 {Intensity interferometry: optical imaging with kilometer
  baselines} {\em Optical and Infrared Interferometry and Imaging V\/} ({\em
  \procspie\/} vol 9907) p 99070M (\textit{Preprint} \eprint{1607.03490})

\bibitem{Ong:2017ihp}
Ong R~A 2017 {Cherenkov Telescope Array: The Next Generation Gamma-ray
  Observatory} {\em {Proceedings, 35th International Cosmic Ray Conference
  (ICRC 2017): Bexco, Busan, Korea, July 12-20, 2017}\/} (\textit{Preprint}
  \eprint{1709.05434})
  \urlprefix\url{https://inspirehep.net/record/1624131/files/arXiv:1709.05434.pdf}

\bibitem{Vincent:2016sjq}
Vincent F~H, Gourgoulhon E, Herdeiro C and Radu E 2016 {\em Phys. Rev.\/} {\bf
  D94} 084045 (\textit{Preprint} \eprint{1606.04246})

\end{thebibliography}
\bibliographystyle{iopart-num}

\end{document}